\documentclass[sigconf, authorversion]{acmart}

\usepackage{amsmath,amsfonts}
\usepackage{acronym}
\usepackage{graphicx}
\usepackage{subcaption}
\usepackage{balance}
\usepackage{todonotes}
\usepackage{array}
\usepackage{hyperref}

\newcolumntype{P}[1]{>{\centering\arraybackslash}p{#1}}

\acrodef{RF}{Radio Frequency}
\acrodef{RFF}{Radio Frequency Fingerprinting}
\acrodef{IoT}{Internet of Things}
\acrodef{DL}{Deep Learning}
\acrodef{CNN}{Convolutional Neural Networks}
\acrodef{SNR}{Signal to Noise Ratio}
\acrodef{BPSK}{Binary Phase Shifting Keying}
\acrodef{SDR}{Software Defined Radio}
\acrodef{RJP}{Relative Jamming Power}
\acrodef{ROC}{Receiver Operating Characteristic}
\acrodef{AUC}{Area Under the Curve}
\acrodef{DL}{Deep Learning}
\acrodef{ML}{Machine Learning}
\acrodef{CSI}{Channel State Information}
\acrodef{CFO}{Carrier Frequency Offset}
\acrodef{BER}{Bit-Error Rate}
\acrodef{TPR}{True Positive Ratio}
\acrodef{FPR}{False Positive Ratio}

\newcommand{\sol}{FingerJam}
\newcommand{\chgclr}{black}

\setcopyright{acmcopyright}
\copyrightyear{2025}
\acmYear{2025}
\setcopyright{cc}
\setcctype{CC-BY}
\acmConference[AsiaCCS '25]{The 20th ACM ASIA Conference on Computer and Communications Security (AsiaCCS)}{August 25-29, 2025}{Hanoi, Vietnam}
\acmBooktitle{The 20th ACM ASIA Conference on Computer and Communications Security (AsiaCCS), August 25-29, 2025, Hanoi, Vietnam}\acmDOI{10.1145/XXXXX.XXXXX}

\begin{document}

\title{Preventing Radio Fingerprinting through Low-Power Jamming}

\author{Muhammad Irfan}
\affiliation{%
\institution{Division of Information and Computing Technology, College of Science and Engineering, Hamad Bin Khalifa University, Qatar Foundtion}
  \city{Doha}
  \country{Qatar}}
\email{muir45306@hbku.edu.qa}

\author{Savio Sciancalepore}
\affiliation{%
\institution{Eindhoven University of Technology}
  \city{Eindhoven}
  \country{Netherlands}}
\email{s.sciancalepore@tue.nl}

\author{Gabriele Oligeri}
\affiliation{%
\institution{Division of Information and Computing Technology, College of Science and Engineering, Hamad Bin Khalifa University, Qatar Foundtion}
  \city{Doha}
  \country{Qatar}}
\email{goligeri@hbku.edu.qa}



\thispagestyle{plain}
\pagestyle{plain}

\begin{abstract}
    Radio Frequency fingerprinting enables a passive receiver to recognize and authenticate a transmitter without the need for cryptographic tools. Authentication is achieved by isolating specific features of the transmitted signal that are unique to the transmitter's hardware. Much research has focused on improving the effectiveness and efficiency of radio frequency fingerprinting to maximize its performance in various scenarios and conditions, while little research examined how to protect devices from being subject to radio fingerprinting in the wild. 
    
    In this paper, we explore a novel point of view. We examine the threat posed by radio frequency fingerprinting, which facilitates the unauthorized identification of wireless devices in the field by malicious entities. We also suggest a method to sanitize the transmitted signal of its fingerprint using a low-power jammer, deployed on purpose to improve devices' anonymity on the channel while still guaranteeing the link's quality of service. Our experimental results and subsequent analysis demonstrate that a low-power jammer can effectively block a malicious eavesdropper from identifying a device without affecting the quality of the wireless link, thereby restoring the privacy of the user when accessing the radio spectrum.
\end{abstract}

\settopmatter{printfolios=true}
\maketitle

\section{Introduction}
\label{sec:introduction}
\ac{RFF} is an emerging physical-layer technique in the context of wireless security and signal intelligence~\cite{jagannath2022comprehensive}. It is based on the finding that every radio transmitter injects a distinctive signature or \emph{fingerprint} into the signal it transmits due to its unique physical characteristics and imperfections~\cite{soltanieh2020_jrfid}. This fingerprint is represented by small variations or imperfections in the transmitter hardware, for example, in its oscillator, power amplifier, or even in the modulation process, which introduce unintentional but characteristic deviations from the ideal signal~\cite{alhazbi2023dayaftertomorrow}. Unlike conventional methods that rely on cryptographic tools, \ac{RFF} takes advantage of the physical layer properties of the signal~\cite{sadighian2024ccnc}, such as \ac{CFO}, phase noise and amplitude fluctuations, among others. Since these characteristics are inherently resistant to masking and cannot be easily replicated, \ac{RFF} can be considered a robust tool for device authentication and identification, even against sophisticated spoofing attacks~\cite{danev2010_wisec}. In fact, \ac{RFF} has the potential to enforce security in communication networks by detecting unauthorized devices, such as rogue transmitters. Furthermore, \ac{RFF} can be critical in the domain of \ac{IoT}, where the sheer number of devices requires efficient and secure methods to ensure that only legitimate devices communicate over networks. Traditional signal processing techniques are not suitable to capture and detect the unique characteristics of \ac{RFF}. However, \ac{DL} has been proven to be able to automatically learn these features directly from the data~\cite{alhazbi2023challenges}. In particular, \ac{CNN} recently turned out to be particularly effective in extracting and learning the features required for \ac{RFF}, either directly or by preprocessing raw signal samples into images~\cite{alhazbi2023ccnc,oligeri2023tifs}.

Although \ac{RFF} is gaining traction in the wireless research community, becoming more efficient and effective, we point out its increasing potential to be used in ways that infringe individual privacy~\cite{abanto2020_macs}. In fact, \ac{RF} devices continuously emit RF signals, which can be eavesdropped and used to train a model characterizing the device. In principle, a passive adversary could collect these emissions, and leverage the associated \ac{DL} model to subsequently identify the device and its owner, thus leading to detailed profiling of individuals' locations and behaviors without their consent. Therefore, \ac{RFF} can potentially be used as an attack vector to identify people and their behavior through the wireless devices they carry, such as smartphones and wearables~\cite{abanto2020_macs,alhazbi2023challenges,ardoin2025,oligeri2024hideprint,smailes2023_ccs,oligeri2023tifs}. Furthermore, if the \ac{RFF} data is collected massively, it could be used to build profiles of technology usage patterns. When combined with other data sources, these profiles become more effective, allowing a more in-depth categorization of individuals. Thus, the ability to uniquely identify devices based on their RF emissions can lead to unintended or unauthorized surveillance and monitoring activities, raising important ethical and privacy concerns.
We emphasize that \ac{RFF} should be considered to the same extent as DNA, since its features are unique. Thus, they cannot be leaked even once, as there is no possibility for the devices to revoke or change them---thus involving the need for such features to be stored and transmitted in a secure way.
\ac{RFF} represents an increasing threat to privacy, as people do not know that RF emissions from their devices can be used to fingerprint them, the same way as it happens for online presence leaking information about the user's location \cite{tomasin2021location}, \cite{caprolu2023watch}, \cite{song2020analyzing}. 
Balancing the benefits of \ac{RFF} with the need to protect individual privacy rights is one of the significant challenges this technology is facing as it becomes an increasingly effective alternative to device authentication.

There have been many attempts to forge radio fingerprints to successfully carry out spoofing attacks, such as the contribution by Danev et al. in~\cite{danev2010_wisec}. However, these works assume that the transmitter is capable of biasing the signal, introducing a fake fingerprint, thus ensuring that the receiver takes the wrong identification of the transmitter. This family of solutions has two issues~\cite{alhazbi2023challenges}: (i) forging a fingerprint requires a transmitter with a high-resolution digital-to-analog converter---this represents a technological barrier, and (ii) the forged fingerprint should take into account the unpredictable effects of the multipath introduced by the wireless channel, which inherently reshapes the signal over-the-air.
In contrast to previous work, we build on the consideration that \emph{challenging} channel conditions can significantly degrade the robustness of \ac{RFF} techniques without affecting the quality of the communication link. We leverage this phenomenon to our advantage and intentionally degrade the quality of the transmitted signal by exploiting another transmitter, i.e., a low-power jammer.

Jamming is a well-known malicious (and deliberate) technique used to prevent reliable delivery of radio messages by injecting a powerful signal at the same frequency used by the target communication~\cite{alhazbi2023ccnc}. The legitimate transmission and the jamming signal collide with the receiver, preventing the correct decoding and demodulation of the transmitted message. Jamming usually finds its main application in the military scenario in disrupting enemy communication and navigation systems~\cite{sciancalepore2023jamming}. Other application scenarios involve jamming devices to block surveillance devices, such as GPS trackers or hidden microphones and cameras. However, to the best of our knowledge, none of the previous contributions investigated the feasibility of jamming to enhance the anonymity and privacy of RF devices against tracking and identification via RFF.

{\bf Contribution.} This work presents \sol, a novel approach to improving privacy in wireless communications by using a controlled, low-power jammer. \sol\ uses jamming to alter the perception of the features of an RF device at the receiver and sanitize radio fingerprints, thus preventing adversaries from identifying specific devices based on their unique transmission characteristics. As additional distinctive characteristics, \sol\ does not require synchronization nor coordination with the device to be protected, emerging as a general-purpose and non-invasive approach to enhance devices' anonymity on the wireless channel. The contribution of this work is manifold: (i) we propose \sol, a low-power jamming technique that introduces controlled interference to obfuscate RF fingerprints without affecting the quality of the legitimate communication link; (ii) we define and employ two metrics, namely k-anonymity and T-anonymity, to quantify the effectiveness of our solution, ensuring that legitimate transmitters are indistinguishable among a set of other devices while achieving similar \ac{BER} of the unjammed communication link; (iii) we establish a defense strategy model demonstrating the efficacy of \sol\ against attackers using state-of-the-art machine learning classifiers, i.e., \acp{CNN} and autoencoders; (iv) our experiments validate the effectiveness of the proposed sanitization technique across different communication links (wired and wireless), showing its capability to prevent identification while maintaining data integrity; \textcolor{\chgclr}{and finally; (v) through real-world experiments, we demonstrate that, as opposed to k-anonymity, T-anonymity is preserved also when the adversary has the possibility to re-train the classifier with data obfuscated by \sol.}

{\bf Paper organization.} The rest of the paper is organized into the following sections. Section~\ref{sec:signal_intelligence} introduces the background on signals used for communication and the \ac{DL} techniques adopted in this study; Section~\ref{sec:scenario} addresses the reference scenario and adversarial model. Section~\ref{sec:measurement} sheds light on the hardware and software tools utilized during this study.  
Sections~\ref{sec:cable_statistic_analysis} and \ref{sec:radio} present the key findings for both cable and radio links, respectively. Section~\ref{sec:related_work} discusses the related work, and, finally, the conclusion is drawn in Section~\ref{sec:conclusion}.

\section{Background on Signal Intelligence}
\label{sec:signal_intelligence}

In this section, we introduce preliminary concepts that we will adopt throughout the paper, i.e., the considered signal representation and \ac{DL} techniques used for \ac{RFF}.

{\bf Signal representation.} A transmitted signal $x(t)$ is characterized by three components, i.e., the amplitude $A$, the phase $\phi$, and the carrier frequency $f_c$ according to Eq.~\ref{eq:bpsk}. Over the years, several techniques have been developed to code information by modulating the amplitude ($A$), the frequency ($f_0)$, or the phase ($\phi$). A smarter way to represent modulated signals is the complex representation of Eq.~\ref{eq:bpsk_IQ}, where both the amplitude ($A$) and the phase ($\phi$) are re-arranged into two orthogonal components, i.e., the {\em in-phase} $I(t) = \cos{(2\pi f_c t)}$ and the {\em quadrature} $Q \sin{(2\pi f_c t)}$. 
\begin{align}
    x(t) & = A \cos{(2\pi f_c t + \phi)}, \label{eq:bpsk} \\
         & = I \cos{(2\pi f_c t)} + Q \sin{(2\pi f_c t)} \label{eq:bpsk_IQ}   
\end{align}
Without loss of generality, in the remainder of this paper, we consider the specific case of \ac{BPSK} modulation, thus setting $Q=0$ and $I = \pm 1$ as a function of the value of the bit to be transmitted.

Starting from a specific bit sequence, the transmitter generates the corresponding I-Q samples and transmits them over the communication link (wired or radio, depending on the scenario). Any receiver, legitimate or not (eavesdropper), collects such I-Q samples after they pass through the link, where they undergo distortion effects based on the experienced channel conditions. Therefore, while transmitted I-Q samples are real values in the form of $[1, 0]$ and $[-1, 0]$ (as a function of the value of the bit to be transmitted), the received ones are affected by a combination of effects due to radio imperfections (the ones used for device fingerprinting), multipath and attenuation. The receiver/eavesdropper is interested in detecting and extracting the fingerprint of the device from such noisy I-Q samples. In the remainder of this paper, we will show how to hide such features without affecting the quality of the link.

{\bf Deep learning.} There has been a significant amount of work to reliably detect transmitter fingerprints and design solutions that take advantage of them efficiently (see Sect.~\ref{sec:related_work}). The leading approach is to resort to \ac{DL} and, in particular, to both \ac{CNN} and {\em autoencoders}. The process involves two steps: first, a \ac{DL} model is trained using batches of I-Q samples, and then, sequences of I-Q samples collected at runtime are tested against the generated model to find out the specific transmitter that generated them. In the remainder of this paper, in line with the relevant literature on RFF, and in particular the well-known concept of \emph{transfer learning}~\cite{torrey2010_igi}, we consider different \acp{CNN} as provided by Matlab2023b and shown in Table~\ref{tab:cnn_comparison}, with the aim of comparing them in terms of training overhead and accuracy.
\begin{table}
\centering
\footnotesize
\caption{Comparison of the \ac{CNN} considered in this work in terms of depth, size, no. of parameters, and input size.}
\begin{tabular}{l|c|c|c|c}
\textbf{Network} & \textbf{Depth} & \textbf{\begin{tabular}[c]{@{}c@{}}Size\\ {[}MB{]}\end{tabular}} & \textbf{\begin{tabular}[c]{@{}c@{}}Parameters\\ {[}Milions{]}\end{tabular}} & \textbf{\begin{tabular}[c]{@{}c@{}}Image\\ Input\\ Size\end{tabular}} \\ \hline
AlexNet             & 8   & 227 & 61   & 227x227 \\ \hline
Squeezenet          & 18  & 5.2 & 1.24 & 224x224 \\ \hline
ResNet-18           & 18  & 44  & 11.7 & 224x224 \\ \hline
ResNet-50           & 50  & 96  & 25.6 & 224x224 \\ \hline
ResNet-101          & 101 & 167 & 44.6 & 224x224 \\ \hline
Inception-v3        & 48  & 89  & 23.9 & 299x299 \\ \hline
Inception-ResNet-v2 & 164 & 209 & 55.9 & 299x299 \\ \hline
NASNet-large        & NA  & 332 & 88.9 & 331x331 \\ \hline
\end{tabular}
\label{tab:cnn_comparison}
\end{table}
\acp{CNN} featured by Matlab2023b are not ready out-of-the-box to handle I-Q samples, so several solutions have been proposed over the years to make them suitable for RFF. In fact, \acp{CNN} were originally designed to recognize objects in images, such as animals or plants, which are quite different from I-Q samples. In this paper, we leverage the image-based pre-processing approach recently proposed by the authors in~\cite{oligeri2023tifs, alhazbi2023dayaftertomorrow}, due to the demonstrated robustness of such pre-processing technique to challenging channel conditions and non-ideal phenomena, such as radios power-cycle. Specifically, we convert raw I-Q samples into images and re-adapt the output layers of the \ac{CNN}. Indeed, the output layers should be modified to match the number of transmitters rather than the categories they were originally designed for---the \acp{CNN} provided by MatLab2023b have been pre-trained on the ImageNet database. Additionally, since the considered networks have already been trained for image recognition, they need to be partially re-trained with the images generated from the raw I-Q samples.

While \acp{CNN} are a perfect fit for multiclass classification, i.e., identification of a transmitter in a well-known pool of devices (k-anonymity, see Sect.~\ref{sec:scenario}), they cannot be used for the detection of a specific transmitter in the wild, i.e., one-class classification (T-anonymity, see Sec.~\ref{sec:scenario}). For this purpose, we resort to \emph{autoencoders}, thus mapping the problem of identifying a transmitter in the wild to a classical anomaly detection problem. We consider the same methodology described for the \ac{CNN}, i.e., implementing a pre-processing technique that translates I-Q samples into images, and subsequently adopt the autoencoder to train, validate and test a model. We follow the classical approach of anomaly detection: we train a model with a dataset of images (generated from a training set of I-Q samples), and then we use such a model to reconstruct images from a test set, comparing the reconstructed images with the actual ones by computing the mean square error. We use the MatlabR2023b implementation of autoencoders and adopt the same parameters used for successful transmitter classification in~\cite{oligeri2023tifs}.

\section{Scenario, Adversarial model and Overview}
\label{sec:scenario}
\begin{figure}
    \centering
    \includegraphics[width=\columnwidth]{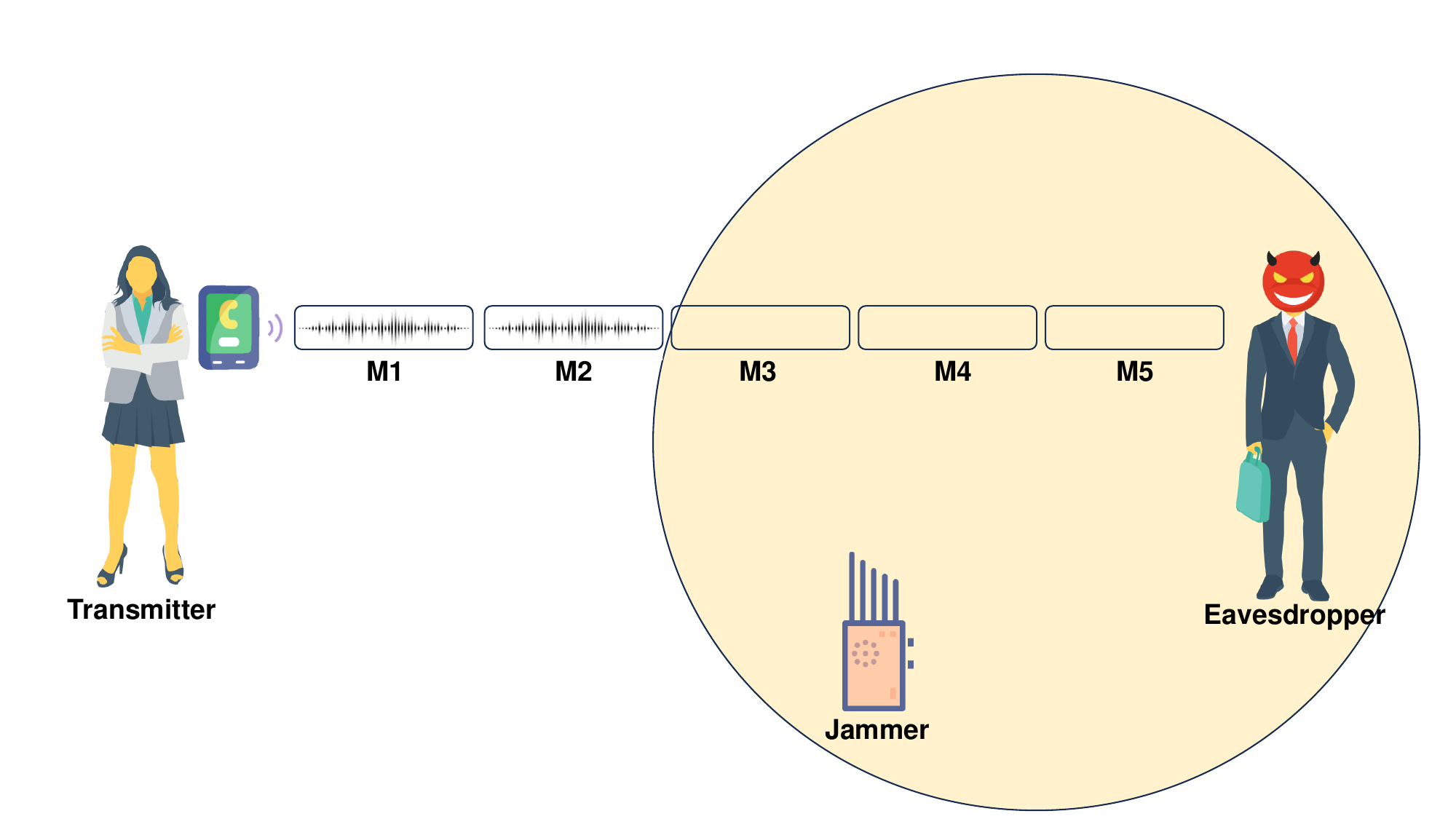}
    \caption{Our scenario, adversary model, and solution. The Eavesdropper wants to identify the presence of the Transmitter at the physical layer, using its RF emissions. The Transmitter protects her identity by deploying a Jammer which removes the radio fingerprint without affecting the quality of the communication. Radio messages M3, M4 and M5 are sanitized of the Transmitter fingerprint. 
    }
    \label{fig:scenario}
\end{figure}

Our scenario and solution consist of three entities: the {\em Transmitter}, the {\em Jammer}, and the {\em Eavesdropper} (also referred to as the {\em Adversary}). We consider a transmitter equipped with a radio device capable of broadcasting radio messages ($M_1, \ldots, M5$) at frequency $f_c$, as shown in Fig.~\ref{fig:scenario}. Each transmitted message is characterized by the transmitter fingerprint, which can be extracted and used by the eavesdropper to infer the presence and identity of the transmitter (transmitter identification). We consider that the eavesdropper already features a \ac{DL} model trained on the transmitter fingerprint. This model can be generated by collecting messages from the transmitter using identifiers from high layers, such as the MAC address of the network interface. We stress that, after the training process has been completed successfully, the identification performed at the physical level does not require packet decoding, and the transmitter identification can be performed directly by acquiring data from the radio spectrum.
The eavesdropper collects raw signals from the radio spectrum and resorts to standard \ac{DL} techniques, such as \acp{CNN} and autoencoders, to identify the presence of the transmitter. 

{\bf \sol\ Overview.} To prevent identification via RFF by the eavesdropper, our solution, namely \sol, requires the transmitter to deploy a jammer that transmits a low-power signal that disrupts the transmitter's fingerprint without affecting the quality of the link. As will be clear in the following, the messages affected by the jammer, i.e., messages $M3, M4$, and $M5$ in Fig.~\ref{fig:scenario}, are sanitized from the transmitter fingerprint and the eavesdropper cannot identify her presence. In this work, we do not make any assumptions about the jammer's position, while we only consider its {\em relative power} to the signal emitted by the transmitter. In fact, the jammer can be deployed as in Fig.~\ref{fig:scenario} to sanitize a specific area far from the transmitter, or it might be featured by the transmitter itself to generate a privacy region in the near-neighboring area of the transmitter. Also, as discussed more in detail in Sec.~\ref{sec:measurement}, we consider a jammer emitting a BPSK-modulated signal (same as the transmitter). As per~\cite{amuru2015}, jamming a BPSK link through another BPSK signal maximizes jamming impact. Overall, \sol\ does not require the jammer to be synchronized with the transmitter, nor to be aware of any features of the transmitter, nor to feature any form of coordination with the transmitter---the transmitter and the jammer can be independent. No modifications are required to the hardware and software running on the transmitter.

Given the previous assumptions, the eavesdropper can deploy the following two strategies to carry out the transmitter identification.

{\bf k-anonymity.} The adversary owns a model with the fingerprint of a pool $k$ of transmitters. The model might have been generated by collecting previous messages from the radio spectrum. In contrast, the pool of $k$ transmitters wants to remain anonymous ({\em k-anonymity}), ideally with a privacy level equal to ($\frac{1}{k}$), i.e., random guess of the transmitter by the eavesdropper. 

{\bf T-anonymity.} The adversary owns a model with the fingerprint of a single (target) transmitter. The model has been generated by processing previously collected messages from such transmitter, and the transmitter aims to avoid identification, i.e., achieving \emph{Target (T)-anonymity}.

We highlight that none of the jamming configurations considered in this paper affects the quality of the legitimate communication link as well as other wireless communications carried out by users in the same area. We take specific care of this concern, and we carefully selected only the configurations that do not change the bit error rate experienced by the receiver, \textcolor{\chgclr}{and so abiding laws about radio spectrum access}. \textcolor{\chgclr}{Overall, low-power jamming as considered in this work aligns with the concept of low-power (intentional) interference, e.g., regulated in US by the Title 47, Chapter 1, Subchapter A, Part 15 of the Code of Federal Regulations titled Radio Frequency Devices~\cite{part15} and, to some extent, the Radio Equipment Directive (RED) in the EU, which sets specific power limits and frequency allocations for RF devices~\cite{mueck2025_red}.}

\section{Measurement set-up}
\label{sec:measurement}
In this section, we detail the hardware and software tools utilized, as well as foundational concepts in RF communication and \ac{DL} that underpin our approach to radio fingerprinting. All the collected data will be publicly released after the acceptance of the paper.

\begin{figure}
    \centering
    \includegraphics[width=\columnwidth]{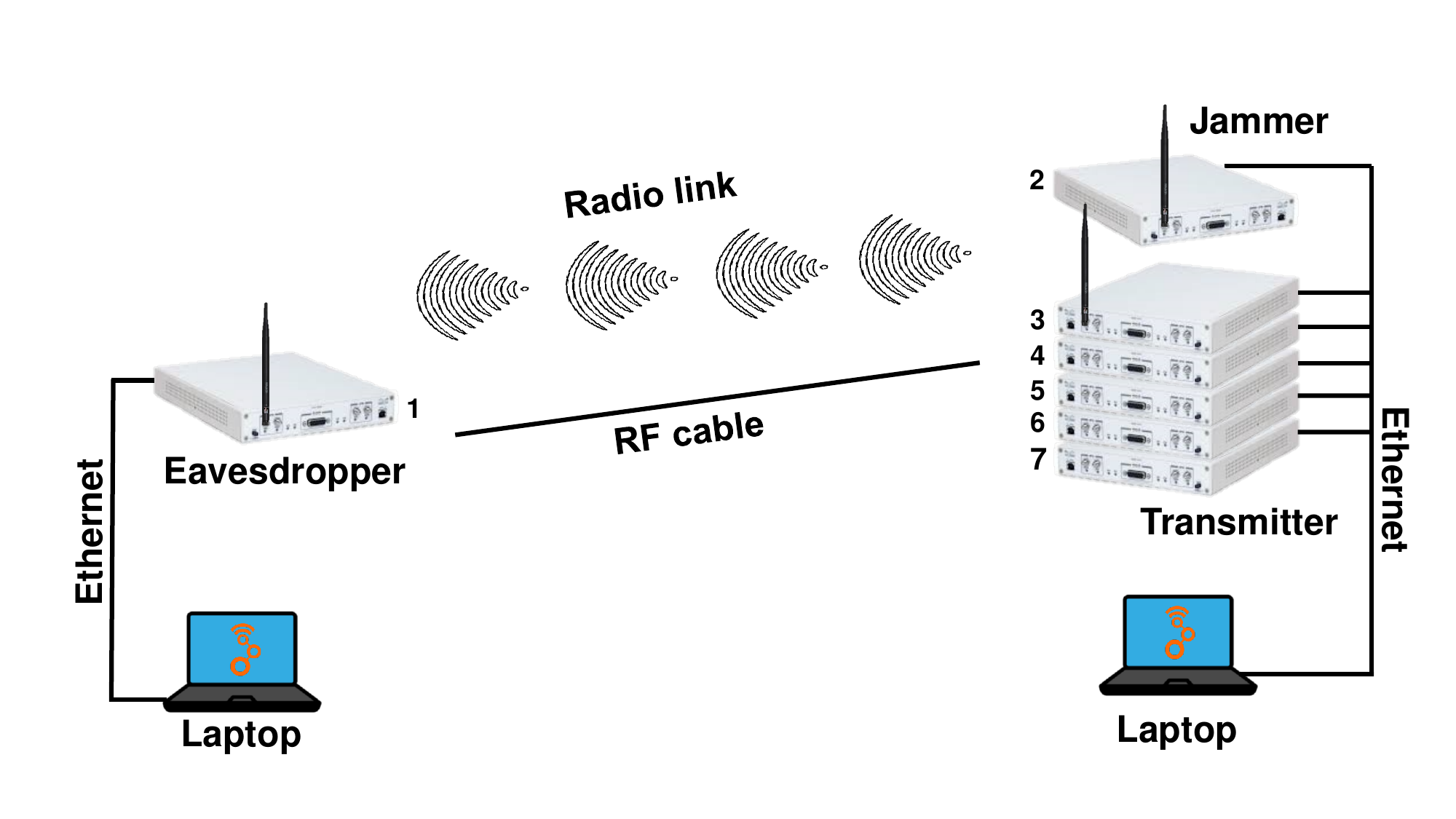}
    \caption{Measurement setup: We considered 7 radios and two types of links, the radio and the cable. Radios 1 and 2 are the Eavesdropper/Receiver and the Jammer, respectively, while Radios 3, 4, 5, 6, and 7 are the Transmitters.}
    \label{fig:hw_setup}
\end{figure}

{\bf Hardware.} Our experimental setup takes into account seven (7) Software Defined Radios (SDRs), specifically USRP X310 models equipped with UBX160 daughterboards and VERT900 antennas. As illustrated in Fig.~\ref{fig:hw_setup}, these radios (referred with identifiers between 1 and 7) are connected to a pair of HP EliteBook laptops, each featuring an i7 processor and 32GB of RAM. Within our experimental measurement campaign, we always consider radio 1 (depicted to the left of Fig.~\ref{fig:hw_setup}) as the eavesdropper and radio 2 as the jammer (shown on top of all radios in Fig.~\ref{fig:hw_setup}). The remaining radios (identifiers 3 to 7) serve as transmitters. We consider both wireless RF and wired connections between the transmitter, the jammer, and the eavesdropper. For all the considered configurations, the jammer features an attenuator with an attenuation factor chosen from one of the following values $\{0, 20, 40\}$ dB. 
We deployed the scenario involving the RF cable by resorting to an RF power splitter/combiner, where the cables coming from the jammer and the transmitter have been connected as input to join the two signals while the output has been connected to the eavesdropper. We use an RG58A/U coaxial cable for all wired link experiments. In the remainder of this work, we consider the {\em Relative Transmission Power} for both the jammer and the transmitters, following the {\em Normalized Gain Value} parameter of the USRP Sink block from GNURadio (see the subsequent description of the software setup). Table~\ref{tab:power_mapping} (Appendix) shows the relation between the value of the Relative Transmission Power and the actual transmission power in dBm.
\begin{table}
\footnotesize
\centering
\caption{Relation between the Relative Transmission Power Value and the actual transmission power of USRP X310.}
\begin{tabular}{c|c}
\hline
\textbf{\begin{tabular}[c]{@{}c@{}}Relative Transmission \\ Power\end{tabular}} & \textbf{Power (dBm)} \\ \hline
0.01    & -9.6     \\ 
0.03    & -9       \\ 
0.04    & -8.5     \\ 
0.05    & -8       \\ 
0.07    & -7.5     \\ 
0.1     & -6.5     \\ 
0.2     & -3.5     \\ 
0.3     & -0.3     \\ 
0.4     &  3.2     \\ 
0.5     &  6.5     \\ 
0.6     &  9.7     \\ 
0.7     & 13.3     \\ 
0.8     & 16.5     \\ 
0.9     & 18.9     \\ 
1       & 20       \\ \hline
\end{tabular}
\label{tab:power_mapping}
\end{table}

In the subsequent sections, we will highlight how (i) the \acp{SDR} are not calibrated, i.e., different radios with the same relative transmission power actually transmit at slightly different power levels, and (ii) we will translate the relative transmission power to the \ac{SNR} experienced at the eavesdropper side, and finally we will show how these parameters affect the fingerprinting process.

We acknowledge that other studies in the RFF domain
often incorporate a larger array of transmitters. We highlight that considering the objective of our research, the usage of a limited number of radios gives additional value to our findings. Indeed, considering the k-anonymity identification strategy, achieving radio anonymity is more difficult when considering a limited set of transmitters rather than a numerous one---with only a few devices, it is harder for one to be confused with another. At the same time, our findings considering T-anonymity are not affected by the number of radios. 

Overall, our research confirms the results of other works using more transmitters in terms of accuracy of the RFF models and also contributes to the foundation for ongoing research in this domain.

{\bf Software.} We consider GNURadio version 3.8 and design an ad-hoc processing chain for the transmitter, the jammer, and the eavesdropper composed of several blocks. The transmitter chain is made up of three blocks: (i) {\em File Source} where we consider a 256-byte string with sequentially increasing values from 0 to 255, (ii) a {\em Constellation modulator}, and finally (iii) a {\em USRP Sink} module. We use a message known at both the transmitter and eavesdropper sides to monitor the quality of the link, thus guaranteeing that all the considered scenarios are not affected by packet loss due to the presence of the jammer (same average values of \ac{BER}). Without loss of generality, we configured the constellation modulator at the transmitter to implement a \ac{BPSK} modulation, and finally, we configured the USRP Sink to transmit on a central carrier frequency of $f_c = 900$ MHz with a sampling rate equal to 2Msps and 512Ksps for the cable and radio scenarios, respectively. Note that all our radio measurements have taken place in an office environment, with people moving around; thus, we chose a lower data rate for the radio link to mitigate out-of-sync phenomena due to the multipath effect caused by people moving in close proximity to the radios. To compensate for these unpredictable phenomena, we considered measurements lasting 60 seconds for the cable and 300 seconds for the radio link. Regarding the eavesdropper chain, we considered five blocks: (i) the {\em USRP Source}, an {\em AGC} block, a {\em Symbol Sync} block, a {\em Costas Loop} block, and finally a {\em File Sync} to store the output of our measurement. The USRP Source has been configured to match the carrier frequency of $f_c = 900$ MHz with a sampling rate identical to the transmitter side; indeed, when considering complex I-Q data (as will become clear in the following), the sampling rate at the eavesdropper side should be greater or equal to the one on the transmitter side, according to the modified Nyquist theorem for complex signals and as also acknowledged by the literature~\cite{sampling_same},~\cite{alhazbi2023dayaftertomorrow}. The AGC block stands for adaptive gain control and has been included to mitigate variations in channel conditions. Finally, symbol sync allows for the recovery of signal timing leveraging the Mueller and M\"{u}eller criterion, and the Costas loop was used to correct for any phase and residual frequency discrepancies.

As for the jammer, we considered a BPSK-modulated signal (same as the transmitter as per \cite{amuru2015}) by considering the block {\em Constellation Modulator} in GnuRadio with 4 samples per symbol and excess bandwidth of 0.035. The jammer has been tuned to the same frequency of the transmitter and receiver, i.e., 900MHz, sample rate, i.e., 512K, while we varied the relative jamming power (RJP) \textcolor{\chgclr}{over our set of experiments}.

\begin{figure}
    \centering
    \includegraphics[width=\columnwidth]{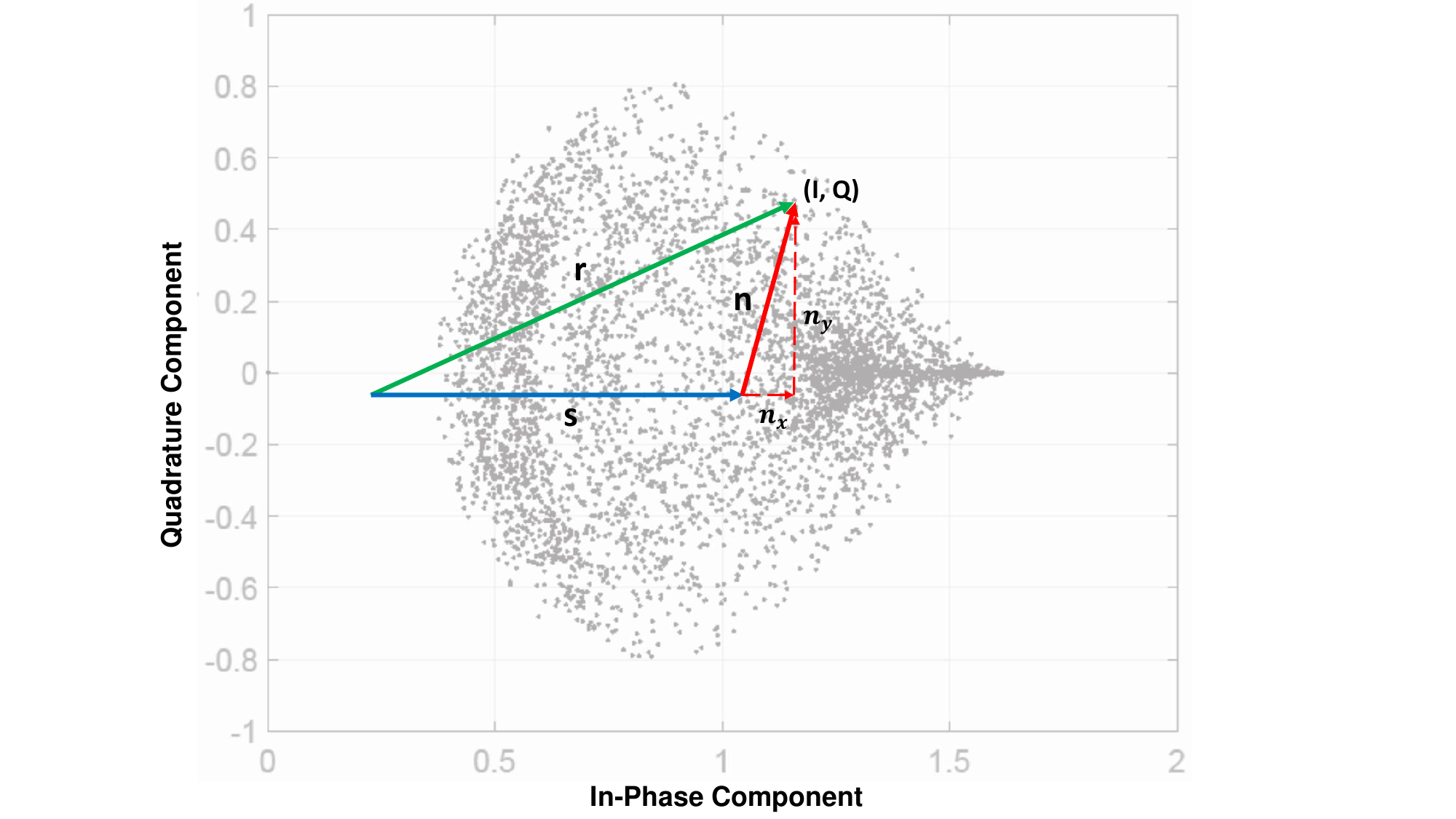}
    \caption{Signal to Noise Ratio (SNR) computation.} 
    \label{fig:snr_computation}
\end{figure}

{\bf Signal to Noise Ratio.} We consider a geometrical definition of the \ac{SNR} as per Fig.~\ref{fig:snr_computation}. The received signal $r$ is made up of two components: the transmitted signal $x$ and the noise $n$. Moreover, assuming \ac{BPSK} as the reference modulation scheme, the quadrature component of the signal $x$ is null. The \ac{SNR} can be calculated as the difference between the power of $r$ and $n$ according to Eq.~\ref{eq:snr}.
\begin{align}
    SNR = P(r) - P(n)
    \label{eq:snr}
\end{align}
While $P(r) = 10 \cdot \log_{10}{(I^2 + Q^2)}$, the power $P(n)$ associated with the noise $n$ can be computed as per Eq.~\ref{eq:noise_power}. 
\begin{equation}
    \label{eq:noise_power}       
    \begin{split}
        P(n) &=  10 \cdot \log_{10}{n^2} \\    
             &=  10 \cdot \log_{10}{(n_x^2 + n_y^2)} \\
             &=  10 \cdot \log_{10}{((I-1)^2 + Q^2)}      
    \end{split}
\end{equation}
Therefore, Eq.~\ref{eq:snr} can be rewritten as Eq.~\ref{eq:snr_final}.
\begin{equation}
    \label{eq:snr_final}   
    \begin{split}
        SNR & = P(r) - P(n) \\
            & = 10 \cdot \log_{10}{(I^2 + Q^2)} - 10 \cdot \log_{10}{((I-1)^2 + Q^2)} \\
            & = 10 \cdot \log_{10}{\frac{I^2 + Q^2}{(I-1)^2 + Q^2}}
    \end{split}
\end{equation}
Thus, the geometric computation of the \ac{SNR} is affected by $Q$ (quadrature component of the received signal $r$) and $n_x = I-1$, i.e., the in-phase component of the noise obtained from the received signal $r$ by subtracting 1.
For the readers' convenience, we summarize the description of the main notation used in this work in Tab.~\ref{tab:notation}.
\begin{table}[H]
\caption{Notation and brief description.
    }
    \label{tab:notation}
    \begin{tabular}{P{1.3cm}|P{6cm}}
       {\bf Notation} & {\bf Description}  \\ \hline
       $x(t)$  & Transmitted signal at time $t$. \\
       $A$ & Amplitude of the transmitted signal. \\
       $f_c$ & Carrier central frequency. \\
       $\phi$ & Phase of the transmitted signal. \\
       $I$ & In-phase component of the signal. \\
       $Q$ & Quadrature component of the signal. \\
       $M_{i}$ & i-th message. \\
       $k$ & Number of transmitters. \\
       $r$ & Received signal. \\
       $P(r)$ & Power of the received signal (in dBm). \\
       $n$ & Noise. \\
       $P(n)$ & Power of the noise (in dBm). \\
       $n_x, n_y$ & Geometrical components of the noise. \\
       $r_A$ & Amplitude of the received signal. \\
       $r_{\phi}$ & Phase of the received signal. \\
    \end{tabular}
\end{table}



\section{Cable Link}
\label{sec:cable_statistic_analysis}

\begin{figure}
    \centering
    \includegraphics[width=\columnwidth]{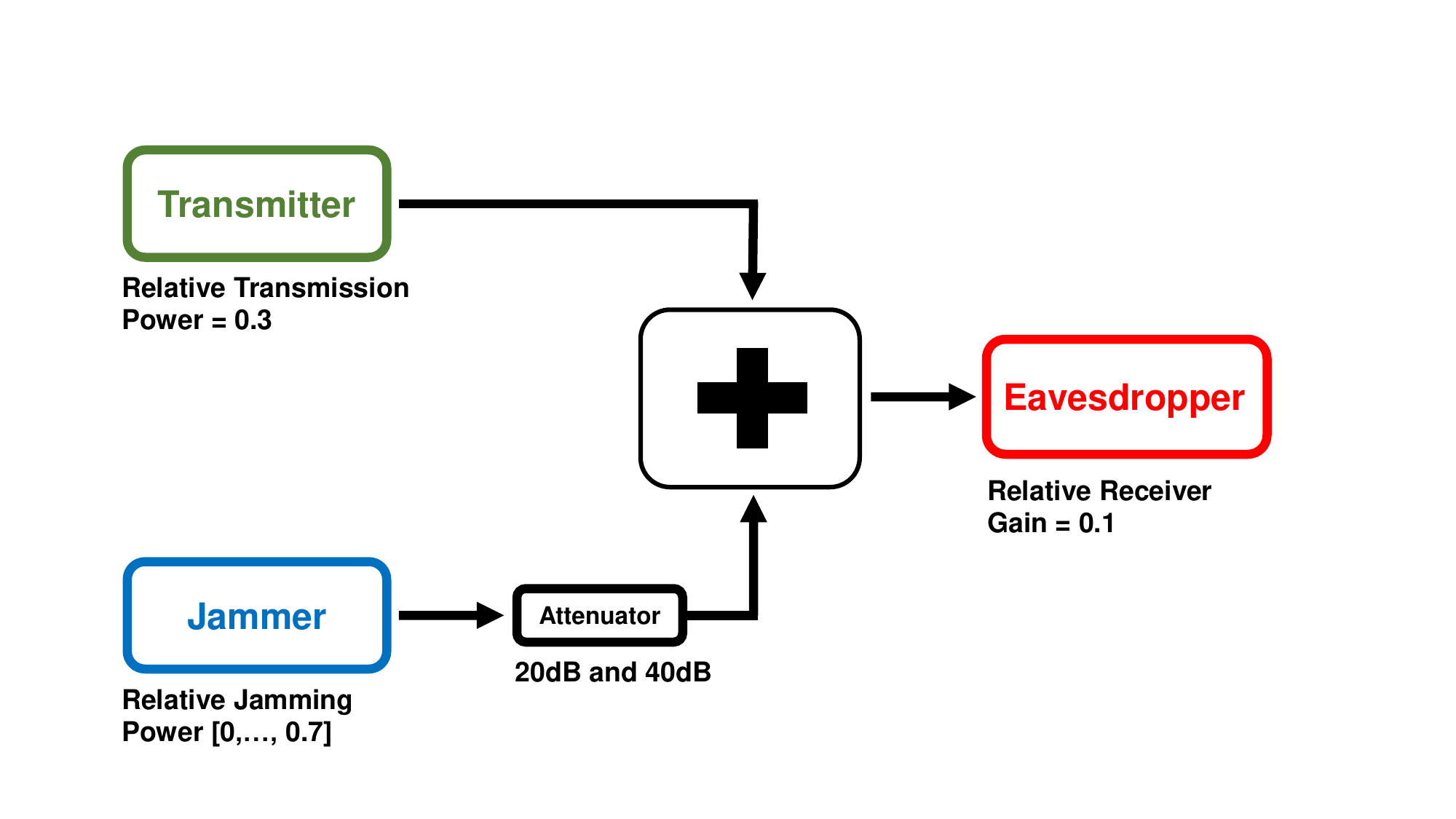}
    \caption{The schematic model of our cable experiments, where the jammer and the transmitter collaborate to prevent the detection of the transmitter by an eavesdropper.} 
    \label{fig:adder}
\end{figure}

In this section, we consider a controlled scenario (low noise) with the five transmitters and the jammer connected to the eavesdropper via a cable. For all experiments, we consider a relative transmission power of 0.3 and a relative receiver gain of 0.1, while we vary the \ac{RJP} between 0 and 0.7 with different ranges as a function of the specific setup. We also consider two attenuators connected to the antenna front-end of the jammer, with attenuation values of 20~dB and 40~dB, respectively. 
Figure~\ref{fig:adder} summarizes our setup.

\subsection{Statistic analysis}
\label{sec:stat_analysis}
In this section, we consider the impact of jamming on the eavesdropper side by only focusing on the physical properties of the signal, i.e., amplitude, phase, and \ac{SNR}. 

\begin{figure}[t]
    \centering
    \includegraphics[width=\columnwidth]{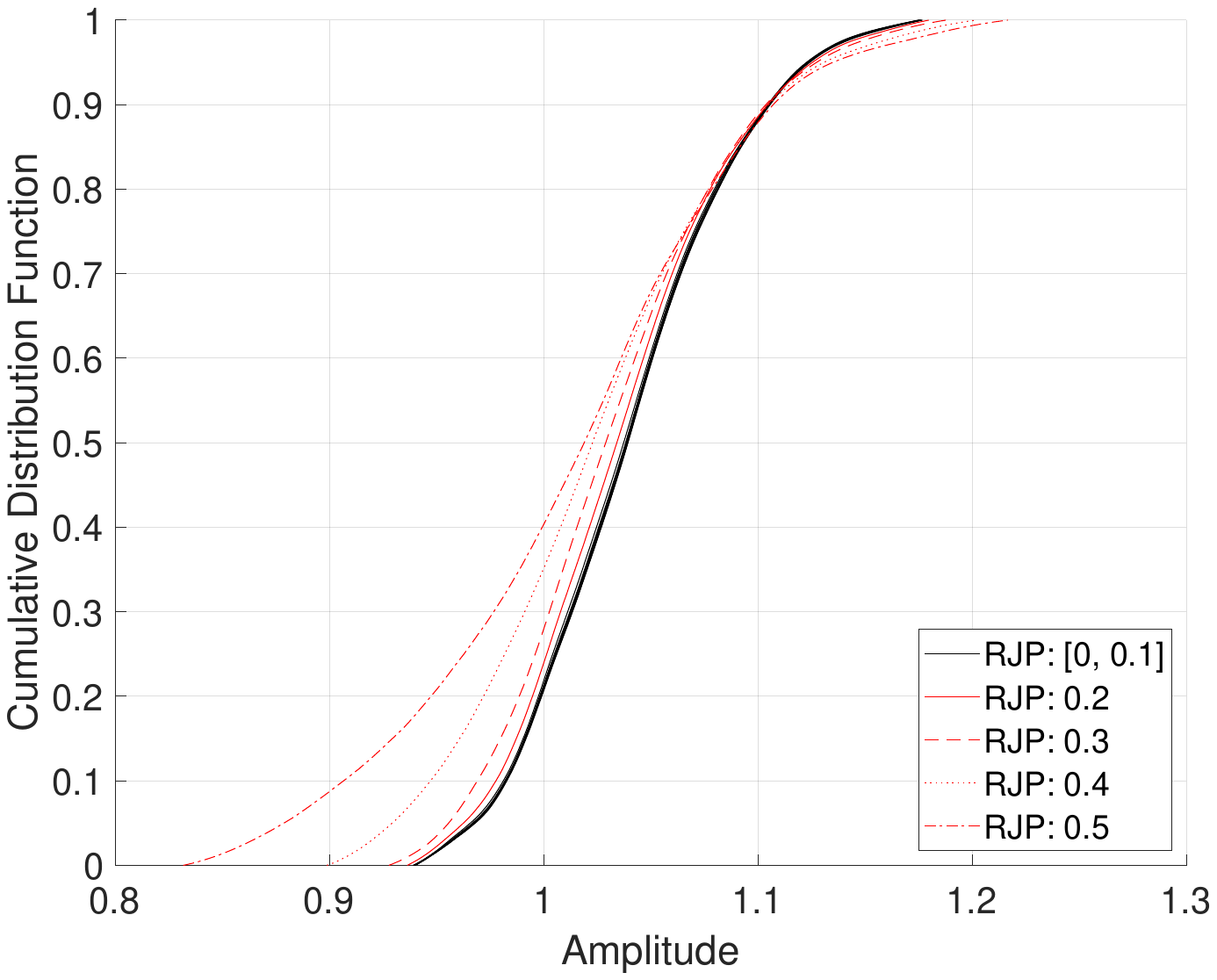}
    \caption{Cumulative distribution function associated with the amplitude of the received signal, varying the RJP between 0 and 0.5. Radios are connected via cables.} 
    \label{fig:cable_amplitude}
\end{figure}

\begin{figure}[t]
    \centering
    \includegraphics[width=\columnwidth]{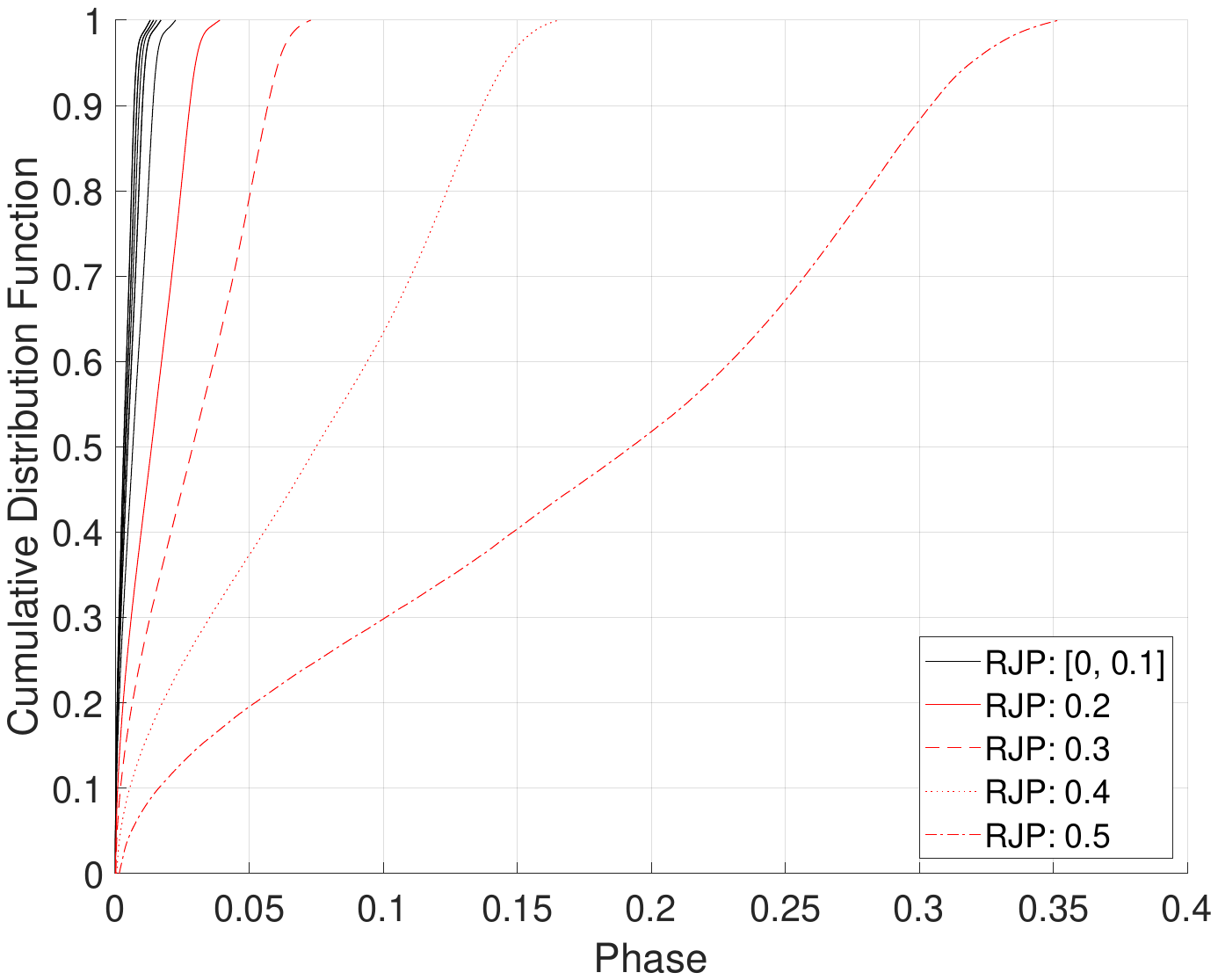}
    \caption{Cumulative distribution function associated with the phase of the received signal, varying the RJP between 0 and 0.5. Radios are connected via cables.} 
    \label{fig:cable_phase}
\end{figure}

Figure~\ref{fig:cable_amplitude} and Fig.~\ref{fig:cable_phase} show the cumulative distribution function associated with the amplitude and phase of the received signal, respectively, when varying the relative jamming power in the set $\{0, 0.03, 0.05, 0.07, 0.1, 0.2, 0.3, 0.4, 0.5\}$ and considering all available transmitters in our setup (average of the group). We compute the amplitude $r_A$ and phase $r_\phi$ of the received signal according to Eq.~\ref{eq:ampli_phase}.
\begin{equation}
    \label{eq:ampli_phase}
    \begin{split}
        r_A &= \sqrt{(I^2 + Q^2)} \\
        r_\phi &= \arctan{\frac{Q}{I}} \\            
    \end{split}
\end{equation}

We observe that for both the amplitude and phase, no significant differences appear when the relative jamming power is in the range $[0, 0.1]$. Furthermore, we observe that the amplitude of the received signal is more robust to the jamming signal with respect to the phase, since significant differences in the amplitude can be noticed only when the relative jamming power is equal to 0.4 and 0.5, respectively. In contrast, even small variations in the relative jamming power (RJP greater than 0.2) affect the phase of the received signal. This can also be confirmed by recalling Fig.~\ref{fig:snr_computation}, introduced previously in Sect.~\ref{sec:measurement}. Even with small relative jamming power values, the received signal keeps the same amplitude ($r \approx x$), but it is affected by a (random) phase rotation ($|n_y|> 0$).

\begin{figure}[t]
    \centering
    \includegraphics[width=\columnwidth]{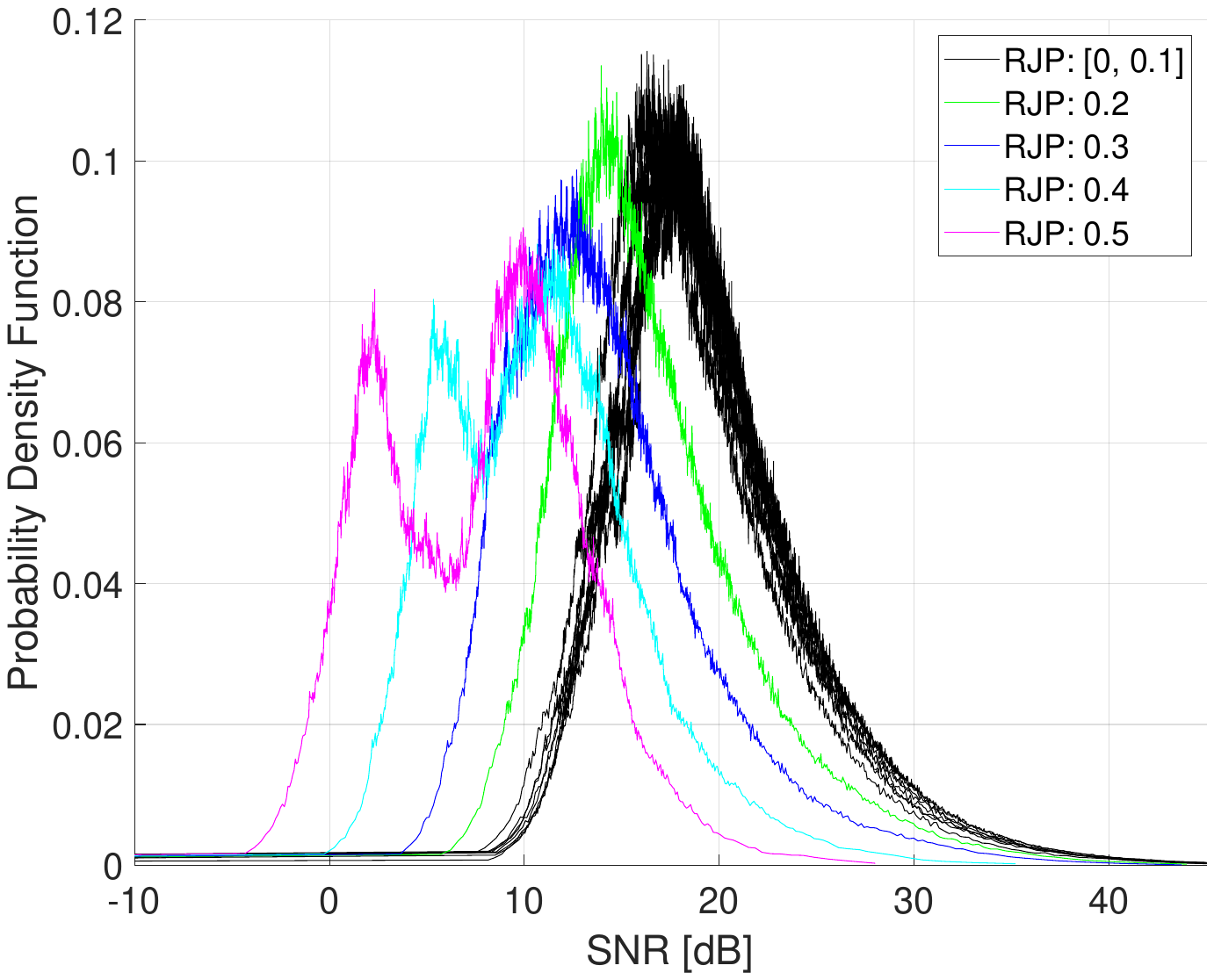}
    \caption{Probability distribution function associated with the signal-to-noise ratio (SNR) of the received signal at the eavesdropper varying the relative jamming power (RJP) between 0 and 0.5. Radios are connected by cables.} 
    \label{fig:cable_snr}
\end{figure}

Furthermore, we consider the \ac{SNR} associated with all the measurements of all the transmitters in our setup, and report the probability density function as shown in Fig.~\ref{fig:cable_snr}. The distribution of the \ac{SNR} values confirms that the SNR profiles at the receiver overlap when the \ac{RJP} is in the range $[0, 0.1]$---as previously observed for amplitude and phase. The effect of jamming on the SNR becomes clearly visible at the eavesdropper side when the \ac{RJP} is greater than 0.2 (solid green line in Fig.~\ref{fig:cable_phase}).

\begin{figure}[t]
    \centering
    \includegraphics[width=\columnwidth]{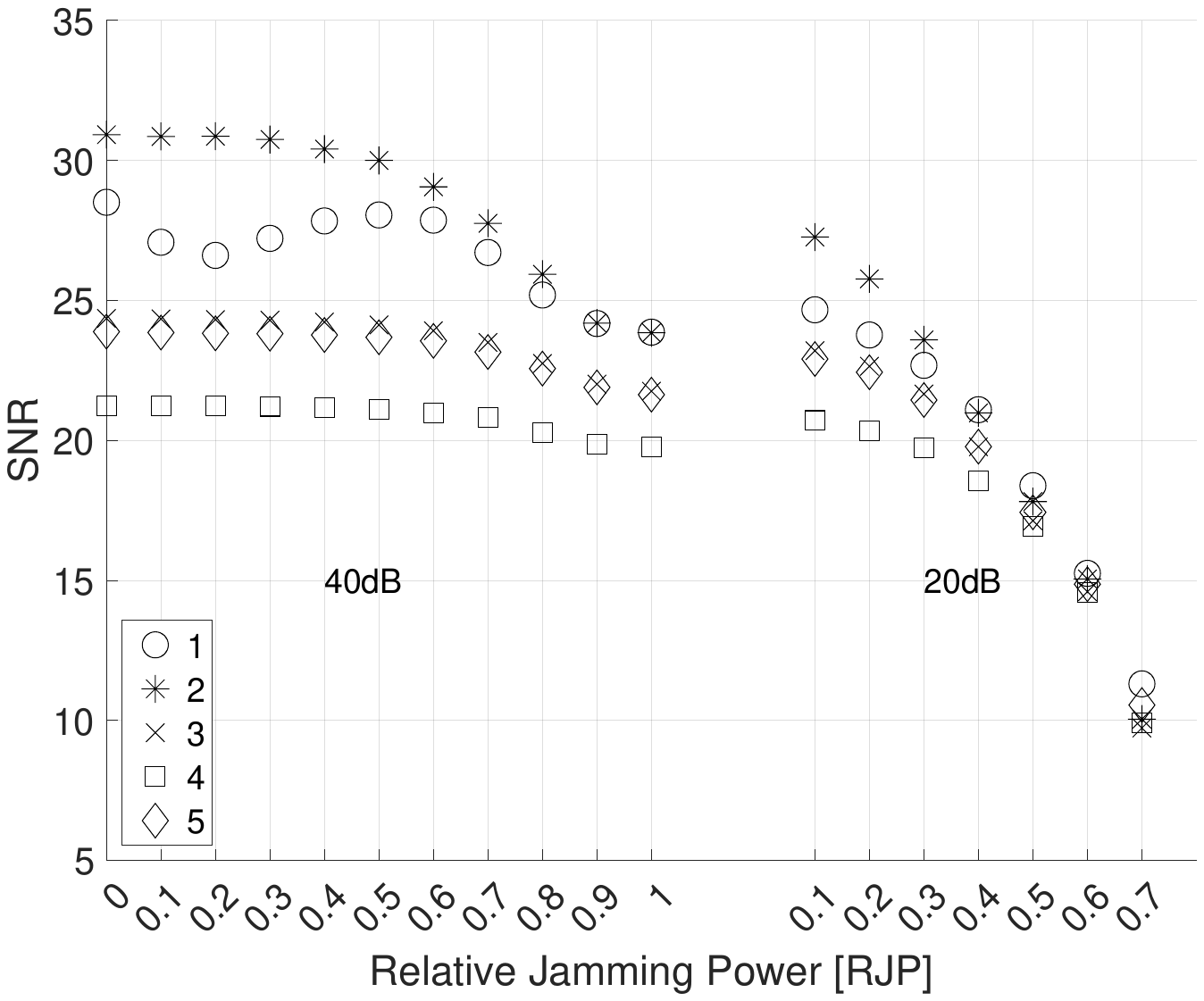}
    \caption{SNR at the eavesdropper side as a function of the RJP, considering different attenuators (20~dB and 40~dB) and all the 5 legitimate transmitters. We set the RJP to 0.3, and the relative receiver gain to 0.1, while the transmitter, the jammer, and the eavesdropper are connected via a cable.} 
    \label{fig:snr_cable_20-40}
\end{figure}

Finally, we focus on the relation between \ac{SNR} and \ac{RJP} when fixing the other parameters of our setup: relative transmission power (set to 0.1) and relative receiver gain (set to 0.3). Figure~\ref{fig:snr_cable_20-40} shows the \ac{SNR} values experienced at the eavesdropper as a function of the \ac{RJP} when considering different attenuators (recall Fig.~\ref{fig:adder}) and each of the five transmitters in our setup.  

First, we observe that the \ac{SNR} decreases when the \ac{RJP} increases, i.e., the noise component on the eavesdropper side increases with respect to the transmitter power. Furthermore, the behavior is affected by the attenuator: when considering the setup including a 40~dB attenuator, the jammer does not affect the \ac{SNR} at the eavesdropper side, as the jamming signal is attenuated and under the noise level of the cable. In contrast, when using a 20~dB attenuator, the SNR decreases from 25~dB (on average) to 10~dB as a function of the \ac{RJP}. We highlight that different transmitters have a different impact on the \ac{SNR} since the transmission power is not calibrated, i.e., when setting the same \ac{RJP}, each transmitter transmits at (slightly) different power. The phenomenon is more evident (up to 10~dB of SNR) with low-power jamming, i.e., with 40~dB attenuation.

{\bf Received signal quality.} Our solution is designed to affect as little as possible the quality of the received signal by minimizing the impact on the bit error rate. Of course, since our solution changes the position of the I-Q samples, it might affect the bit error rate in a very noisy environment. As an example, when the distance between the transmitter and the receiver is sufficiently large, even the smallest interference (jamming signal) can increase the bit error rate. The same phenomenon could be observed in particularly noisy environments, e.g., when many people are in the close proximity of the transmitter-receiver link and strong multipath fading can produce periods where the quality of the link is low. The analysis performed in this section aims to prove that the quality of the signal remains high when the (normalized) transmission power of the transmitter and the jammer are similar, i.e., 0.1 and 0.2, respectively. Indeed, the analysis of the amplitude (Fig.~\ref{fig:cable_amplitude}), the phase (Fig.~\ref{fig:cable_phase}), and the \ac{SNR} (Fig.~\ref{fig:cable_snr}) confirms our claim. It is worth mentioning that such analytics have been evaluated in the best possible conditions, i.e., a cable connecting the transmitter, the receiver, and the jammer. Indeed, when introducing a radio link between the transmitter and the receiver, the reliability of the fingerprint becomes challenging without adding any external interference, thus the channel itself represents a preliminary (and not reliable) privacy preserving barrier against malicious fingerprinting. Unfortunately, relying on bad radio channel conditions for enforcing the privacy of the transmitter is not \textcolor{\chgclr}{effective}, thus the importance of our solution, which \textcolor{\chgclr}{provides} anonymity in the best possible channel conditions.

\subsection{Radio Frequency Fingerprinting}
\label{sec:rff_cable_results}

In this section, we consider both the \acp{CNN} and the autoencoders as \ac{DL} classifiers for \ac{RFF} in the reference setup of Fig.~\ref{fig:adder}. We discuss the performance of our setup when considering different neural networks, different \ac{RJP} values, and finally, different \ac{SNR} conditions.

\begin{figure}
    \centering
    \includegraphics[width=\columnwidth]{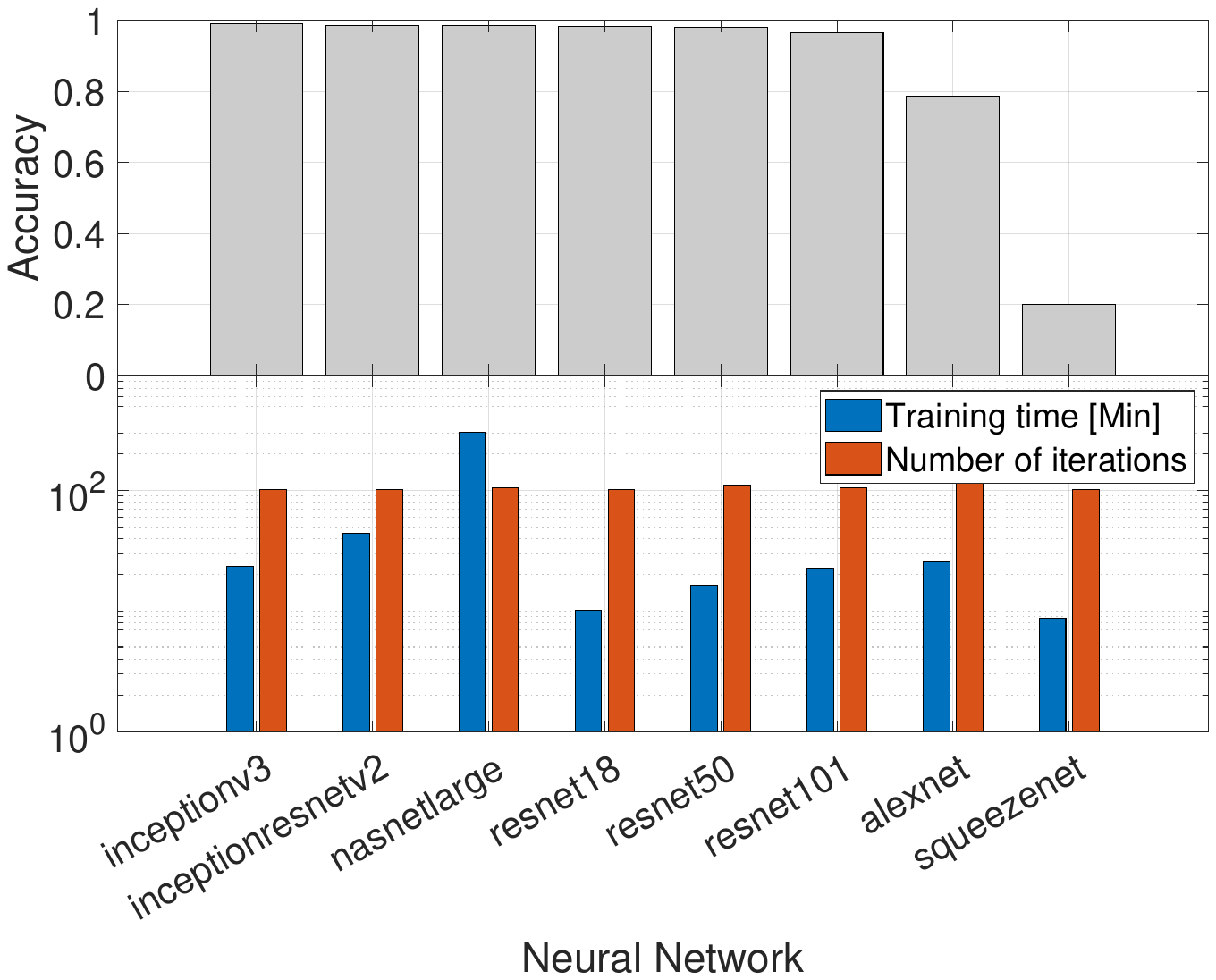}
    \caption{Baseline cable scenario (no jamming): Accuracy as a function of the \ac{CNN} considering 5 transmitters. The bottom figure shows the training time (minutes) and the number of iterations.} 
    \label{fig:no_jamming_cable}
\end{figure}

{\bf k-anonymity.} We start our analysis from a reference scenario where the transmitters are connected via cable to the eavesdropper without a jamming signal. We consider the same parameter configuration introduced above, i.e., we set the relative transmission power to 0.3, and the relative receiver gain to 0.1. In this scenario, the pool of the transmitters ($k=5$) aims to avoid identification, keeping the k-anonymity equal to $\frac{1}{5} = 0.2$. Figure~\ref{fig:no_jamming_cable} shows the performance of eight \acp{CNN} for the RFF task (recall Table~\ref{tab:cnn_comparison}) in terms of accuracy, training time, and number of iterations to achieve the declared accuracy. We observe that all networks, except \emph{squeezenet} and \emph{alexnet}, perform very well in identifying the transmitter. The number of iterations (100) has been set manually, while the training time is a function of the complexity of the network. We observe that---under the scenario considered for this case---the network achieving the best trade-off between accuracy and training time is \emph{resnet18}, reporting an accuracy of 0.98 and a training time of about 10.13~minutes. We stress that, without jamming, an eavesdropper can easily identify the transmitter by resorting to any of the \ac{CNN} considered in our analysis, except {\em alexnet} and {\em squeezenet}.

\begin{figure}
    \centering
    \includegraphics[width=\columnwidth]{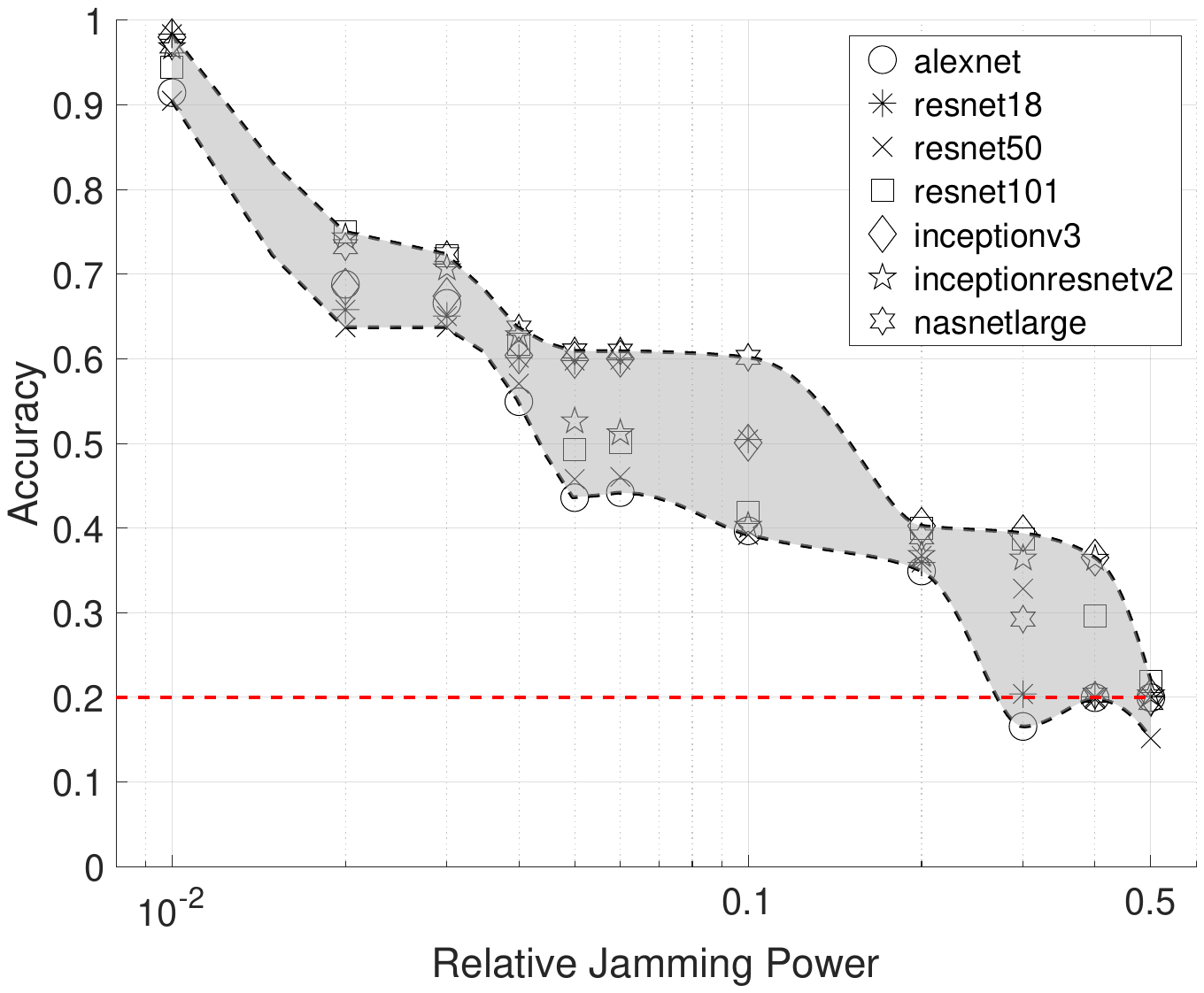}
     \caption{Accuracy as a function of the Relative Jamming Power (RJP) considering 5 transmitters and a jammer (attenuated by 20dB). The dashed red line represents the maximum theoretical uncertainty (random guess), while all measurements are characterized by BER = 0.} 
    \label{fig:jamming_cable}
\end{figure}

We now consider the more general case where the transmitter-eavesdropper link experiences the presence of a jammer. For this experiment, we only consider the 20dB attenuator (recall Fig.~\ref{fig:adder}, and the standard setup shown in Fig.~\ref{fig:hw_setup}, consisting of 5 transmitters, 1 jammer and 1 eavesdropper. Also, note that the BER is 0 for all the considered RJP values. Figure~\ref{fig:jamming_cable} shows the accuracy of the various considered \acp{CNN} as a function of \ac{RJP}. We also depict through a dashed red line the maximum theoretical uncertainty of a classifier (random guess), i.e., $\frac{1}{5}=0.2$.

Figure~\ref{fig:jamming_cable} shows that \acp{CNN} are very sensitive to the jamming phenomena. In fact, when \ac{RJP} equals 0.02, the accuracy drops to 0.7 (on average). Furthermore, we observe that, when \ac{RJP} equals 0.1, the accuracy is between 0.5 and 0.6 as a function of the considered neural network. Networks with higher complexity, e.g., \emph{nasnetlarge} and \emph{inceptionresnetv2}, are much more robust to noise (jamming), but require more time for the training process (recall Fig.~\ref{fig:no_jamming_cable}). Finally, we highlight with a dashed red line the lower bound equal to 0.2 ($5$-anonymity), i.e., the random guess of the transmitter. On average, such a lower bound can be achieved when $RJP=0.5$ while still preserving the quality of the communication link (BER=0).

\begin{figure}
    \centering
    \includegraphics[width=\columnwidth]{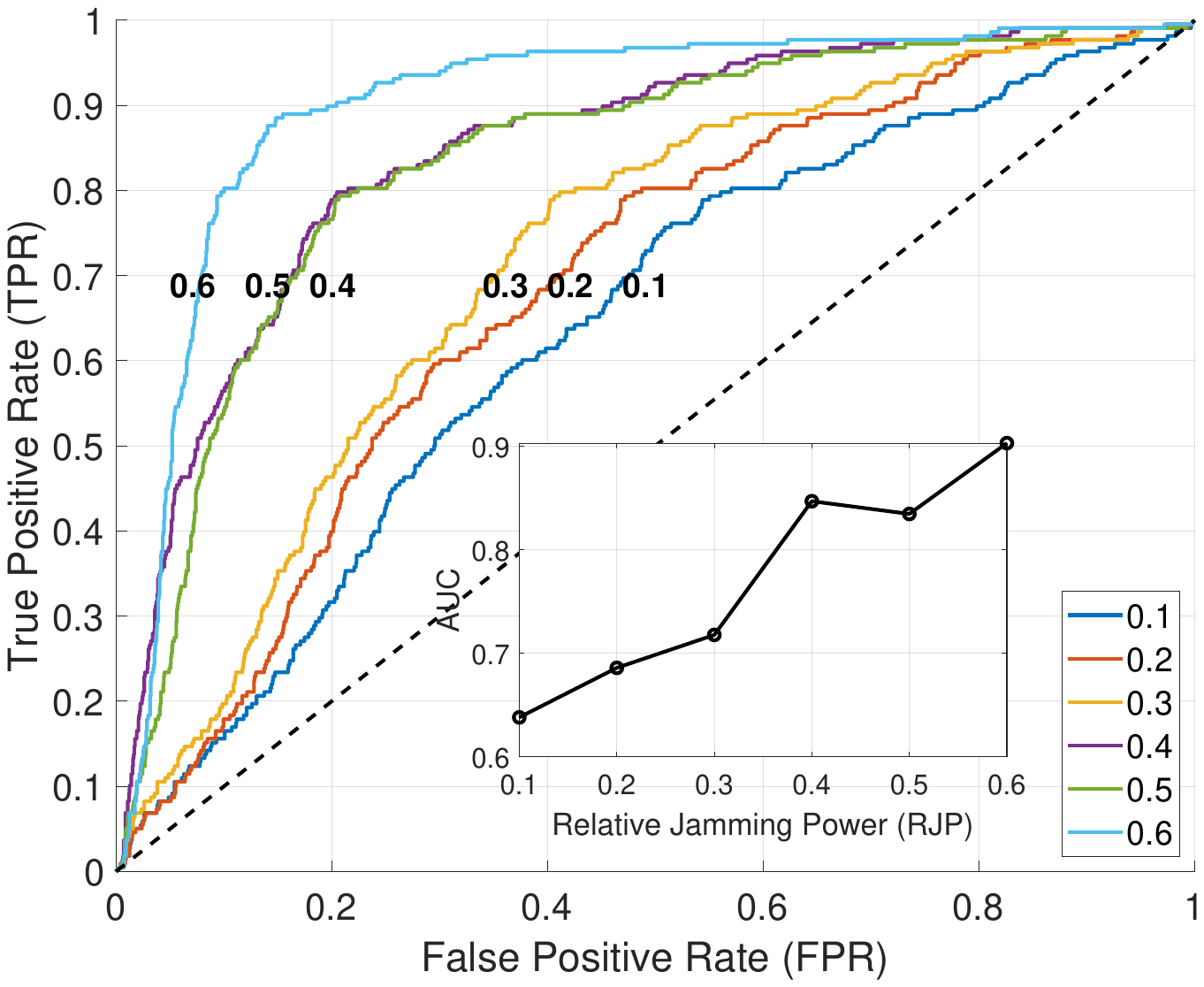}
    \caption{Performance of the autoencoders with a cable link: ROC curve as a function of the \ac{RJP}. The dashed black line represents the random guess, i.e., the classifier cannot distinguish a jammed signal from the unjammed one.} 
    \label{fig:roc_cable_20-40}
\end{figure}

{\bf T-anonymity.} We now consider the scenario in which the adversary focuses on a target transmitter. For this scenario, autoencoders (one-class classification) are the best fit. We highlight that the main objective of this section is to investigate the level of the power that should be injected by the jammer in order to make the received signal different from the actual one. In such a scenario, the bit error rate of the legitimate link is not affected, while the signal at the receiver looks different. Therefore, RFF cannot be used by the adversary to identify the device in an unauthorized way, while preserving the quality of the link. Figure~\ref{fig:roc_cable_20-40} shows the \ac{ROC} curve, i.e., the \ac{TPR} of the classification as a function of the \ac{FPR}, when considering all transmitters of our setup and different \ac{RJP} values, in the range $[0.1, 0.6]$. We labeled the samples (images) not affected by the jammer as the positive class (training set), while testing all transmitters on different \ac{RJP} in the range $[0.1, 0.6]$.
\ac{RJP} = 0.1 is the worst performing (\ac{AUC} $\approx 0.63$): the autoencoder cannot detect the difference between an unjammed signal and a jammed one when the \ac{RJP} = 0.1 and the jamming signal is attenuated by 20dB. Therefore, the fingerprint is still present when considering such a setup: the features extracted during the training (no jamming) are the same as the ones extracted during the testing (in the presence of a weak jammer). When increasing the \ac{RJP}, the performance of the classifier increases; in fact, it becomes easier for the autoencoder to detect the difference between the unjammed signal and the jammed one. In particular, we stopped our analysis at \ac{RJP} equal to 0.6 (AUC $\approx 0.9$), because this is the highest relative jamming power that does not yet affect the bit error rate on the link. For completeness, we report the random guess as the dashed black line in Fig.~\ref{fig:roc_cable_20-40}. We highlight that the dashed black line (random guess) in this figure represents the worst possible performance of the added noise in terms of protection of the device from RFF. In fact, the closer performances are to the dashed black line, the more features at testing time (under jamming) are similar to the ones at training time (no jamming), meaning that the jammer is not able to change them to achieve anonymity. Thus, with reference to the performance curves reported in Fig.~\ref{fig:roc_cable_20-40}, the curves in the top-left part of the figure are the ones characterized by higher performances in terms of anonymity provided to the RF device against RFF. In fact, using such RJP values, the device looks different to the autoencoder, although being the same, due to the deployment of the low-power jammer as per our solution.
Finally, we consider the Area Under the Curve (AUC) for each of the previous ROC curves, and we report the AUC as a function of the relative jamming power in the inset of Fig.~\ref{fig:roc_cable_20-40}. The trend of the AUC confirms that the autoencoders perform better (higher AUC) when the relative jamming power is high (RJP = 0.6), which proves that specific jamming signals affect the reception process of the transmitted signal without changing the quality of the link, i.e., all the transmitted bits have been correctly received in our measurements.

\begin{figure}
    \centering
    \includegraphics[width=\columnwidth]{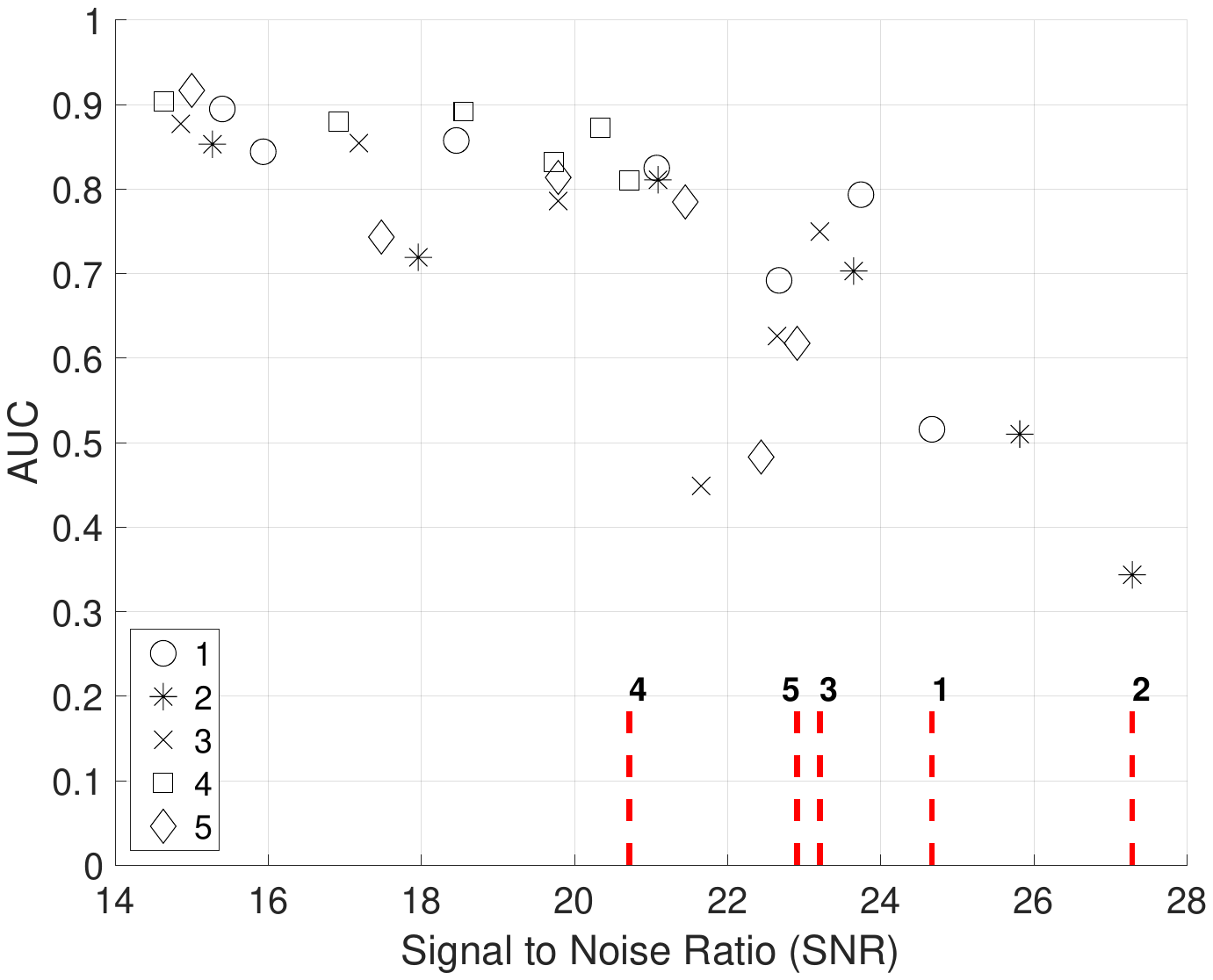}
    \caption{Performance of the autoencoders: AUC as a function of the \ac{SNR}, considering each transmitter one by one.} 
    \label{fig:auc_cable_20-40}
\end{figure}

We now consider the breakdown analysis of each transmitter and compute the \ac{AUC} as a function of the \ac{SNR}, as shown in Fig.~\ref{fig:auc_cable_20-40}. For each measurement configuration (transmitter ID, \ac{RJP}), we computed the \ac{SNR} associated with the measurement and the \ac{AUC} as the indicator of the classifier performance. We note that we select the AUC as our metric rather than the accuracy, as the former is more suitable when considering unbalanced datasets, as in our scenario. 
Figure~\ref{fig:auc_cable_20-40} confirms that the performance of the classifier (autoencoder) decreases when the \ac{SNR} increases (smaller \ac{RJP}). In fact, when the relative jamming power is small, the quality of the communication link is high, so the fingerprint can be easily extracted. In contrast, when the \ac{SNR} is smaller than 20dB, the fingerprint is completely removed from all transmitters, allowing the classifier to distinguish between jammed and unjammed signals (AUC greater than 0.8 on average). Finally, we observe that the transmitters behave in different ways when subjected to jamming. We highlight through a dashed red line in Fig.~\ref{fig:auc_cable_20-40} the largest experienced \ac{SNR} for each transmitter when considering the same measurement configuration and, in particular, the same \ac{RJP} in the range $[0.1, 0.6]$. We explain this phenomenon by recalling Fig.~\ref{fig:cable_snr}: the \acp{SDR} are not calibrated, thus transmitting with different transmission power when setting the same relative transmission power.

\section{Radio Link}
\label{sec:radio}
We also validated our findings in a real wireless link. For this scenario, we recall the measurement setup shown in Fig.~\ref{fig:hw_setup} and consider a radio link among the three entities: the transmitter, the jammer, and the eavesdropper. In our deployment, we consider the transmitter and the eavesdropper to be located a few meters away inside the same (office) room, while we place the jammer 10 meters away outside the room (no Line of Sight). As for the previous experiments, we consider one jammer, one eavesdropper, and five transmitters. Moreover, we set a relative transmission power of 0.8, a relative receiver gain of 0.8 and two values for the relative jamming power: 0 (no jamming) and 1 (jamming at the maximum power). Finally, for the jammer RF front-end, we perform different measurements by considering three different configurations, i.e., No attenuation, 20~dB, and 40~dB. 

\begin{figure}
    \centering
    \includegraphics[width=\columnwidth]{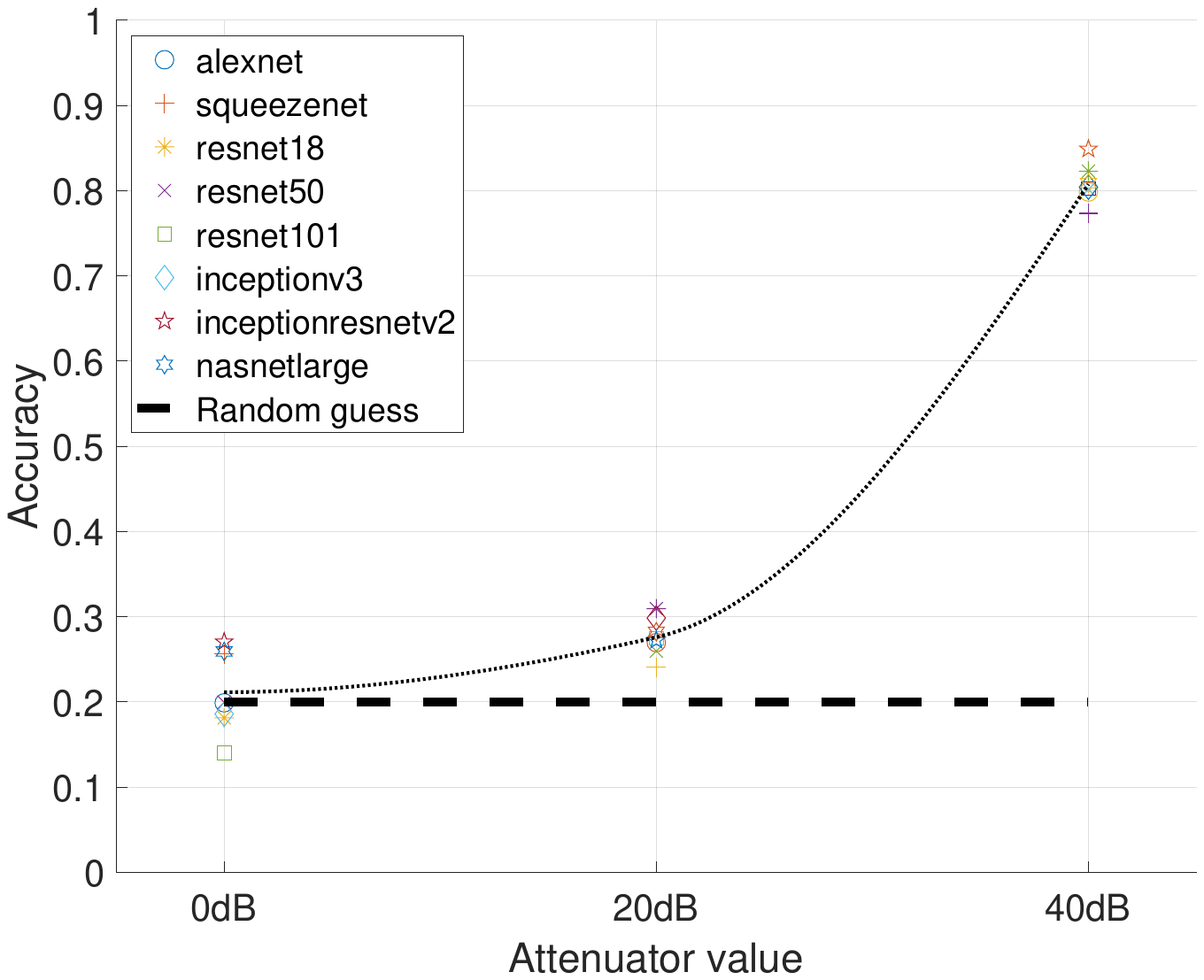}
    \caption{Performance of the \acp{CNN} with a radio link: accuracy as a function of the attenuation at the jammer, with various \acp{CNN}.} 
    \label{fig:radio_cnn}
\end{figure}

{\bf k-anonymity.} We train our model by considering a fraction (60\%) of all measures collected from all transmitters given a specific value of the attenuator, and we use the remaining fraction for validation and testing, 20\% each. Figure~\ref{fig:radio_cnn} shows the performance of the different \acp{CNN} as a function of the value of the attenuator connected to the jammer. The performance of \acp{CNN} is significantly affected by the noise power injected by the jammer, i.e., the value of the attenuator. When there is no attenuation, all networks perform very badly, i.e., the accuracy is about 0.2---the same as a random guess (dashed black line). When the attenuation increases (20~dB and 40~dB), the performance of the \acp{CNN} increases accordingly, up to 0.8---we stress that we ran the measurements in an extremely noisy and dynamic environment, thus being affected by strong real-world multipath phenomena.
Our analysis shows that all the networks have similar accuracy for all the configurations, and the jammer is successful at removing the fingerprint when considering enough \ac{SNR}, i.e., \ac{RJP} equal to 1 and attenuation smaller than 20~dB. Under such assumptions, the best option for the eavesdropper is a random guess of the transmitter.

\begin{figure}[!h]%
    \centering
    \subfloat[\centering No attenuation]{{\includegraphics[width=5.7cm]{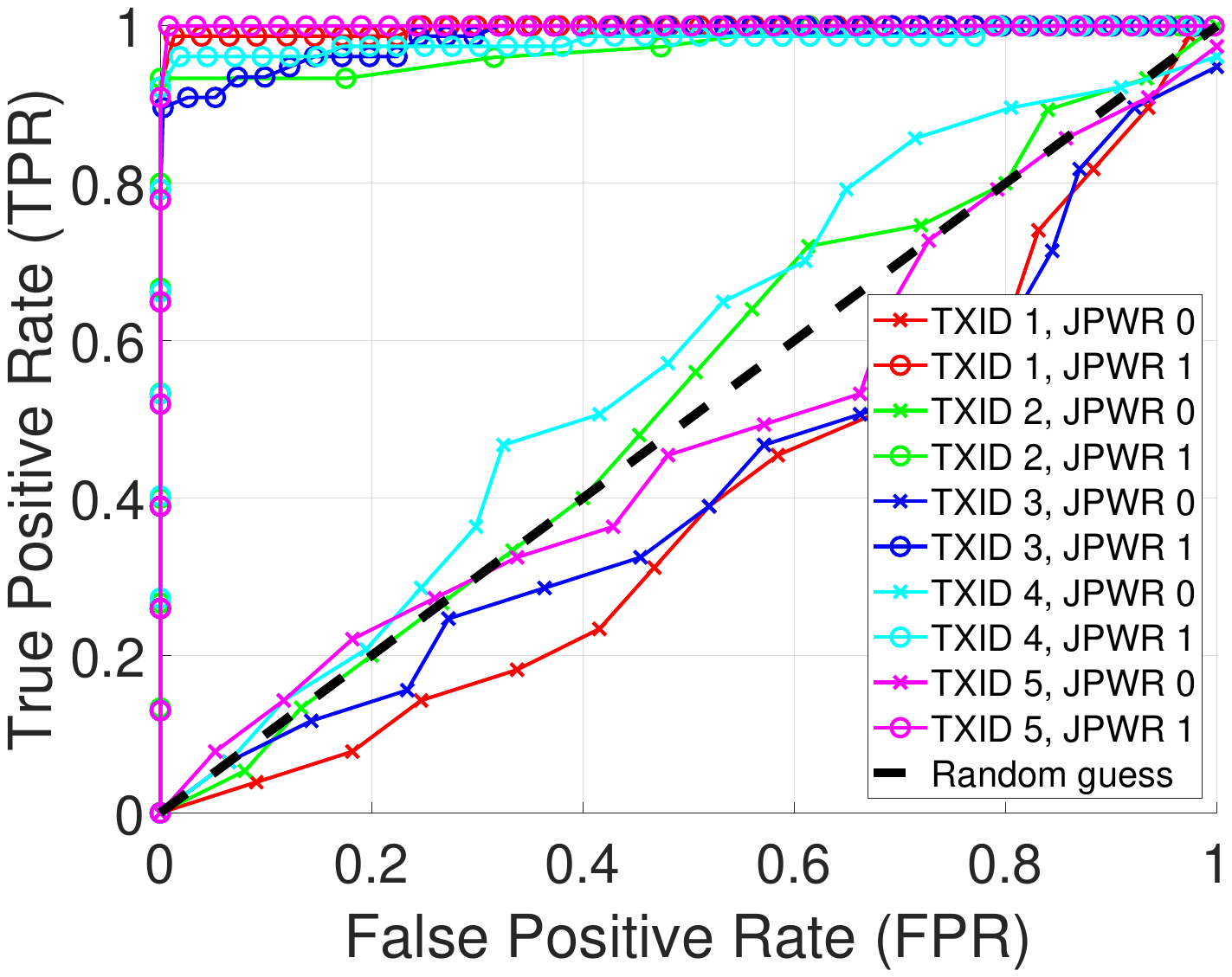}}}%
    \ 
    \subfloat[\centering Attenuator: 20dB]{{\includegraphics[width=5.7cm]{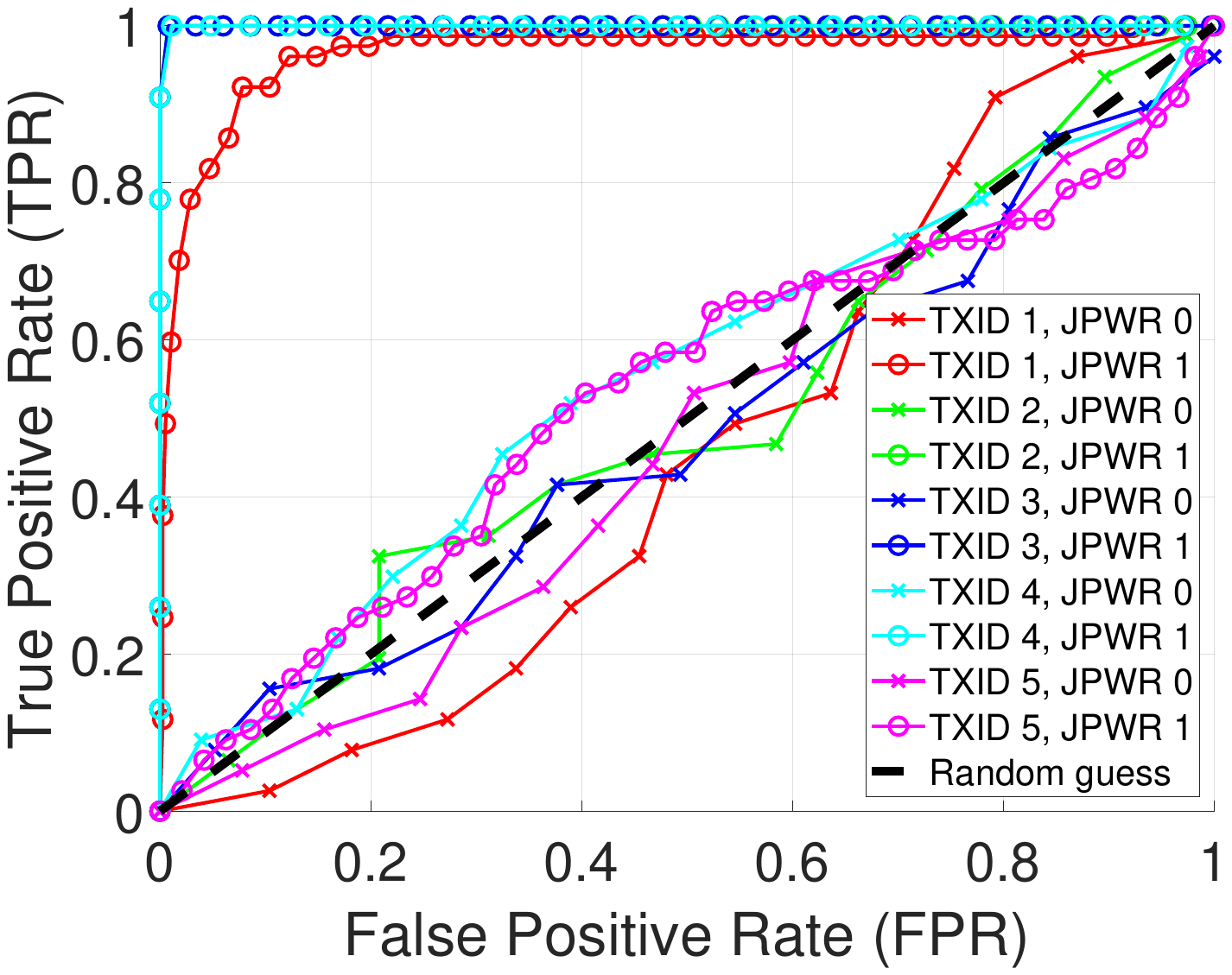}}}%
    \ 
    \subfloat[\centering Attenuator: 40dB]{{\includegraphics[width=5.7cm]{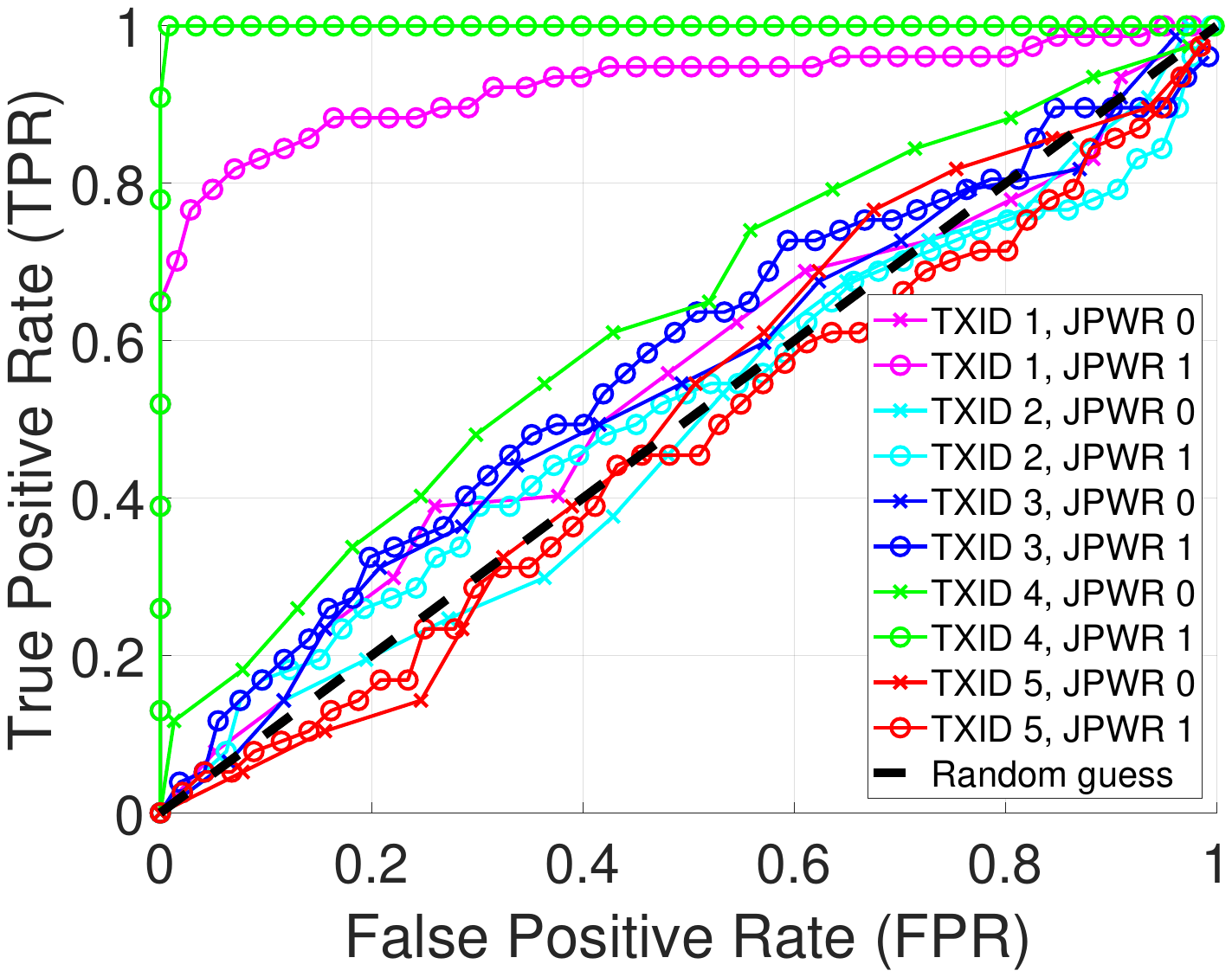}}}%
    \caption{Performance of the autoencoders in a radio link: we consider 5 transmitters, two jamming powers (0 and 1), and three attenuations (no attenuation, 20~dB, 40~dB) for the RF jamming front-end.}%
    \label{fig:roc_radio}%
\end{figure}

{\bf T-anonymity.} We now consider an eavesdropper with a model trained on a specific target device (one of the five transmitters in our pool). We train a classifier constituted by an autoencoder with measures collected with jamming power equal to zero (JPWR 0), and test the same model on measures affected by jamming (JPWR 1) and different attenuation values, i.e., no attenuation, 20~dB and 40~dB. Figure~\ref{fig:roc_radio} shows the performance of the autoencoder in terms of \ac{TPR} as a function of \ac{FPR} (ROC curve). 

When no attenuation is considered on the jammer side (Fig.~\ref{fig:roc_radio}(a)), the autoencoder distinguishes very well between no jamming (JPWR 0) and jamming at the maximum relative transmission power (JPWR 1) for all transmitter IDs (TXID 1 to 5). In fact, the ROC curves associated with JPWR = 0 lie close to the random guess (dashed black line), while the ROC curves associated with JPWR = 1 are characterized by a steep variation when $FPR > 0$. The performance of the autoencoder is affected by the attenuator, i.e., when the jamming signal is weak, as shown in Fig.~\ref{fig:roc_radio}(c). In the latter case, all the ROC curves (except TXID 1 - JPWR 1 and TXID 4 - JPWR 1) have the same trend of the random guess (black dashed line), i.e., a weak jamming signal (attenuated by 40~dB) does not allow the autoencoder to distinguish between a jammed and an unjammed signal at the receiver side. Finally, we highlight that for each transmitter, we considered both JPWR = 0 and JPWR = 1, so to validate our results. From these results, we can conclude that, in a real-world radio link, low-power jamming protects the anonymity of RF devices without affecting the quality of the communication link. However, the necessary jamming power should be calibrated on a case-by-case basis, depending on the specific device to be protected.

\begin{figure}[!h]
    \centering
    \begin{subfigure}{\columnwidth}
        \includegraphics[width=\columnwidth, trim={10mm 0mm 10mm 0}, clip]{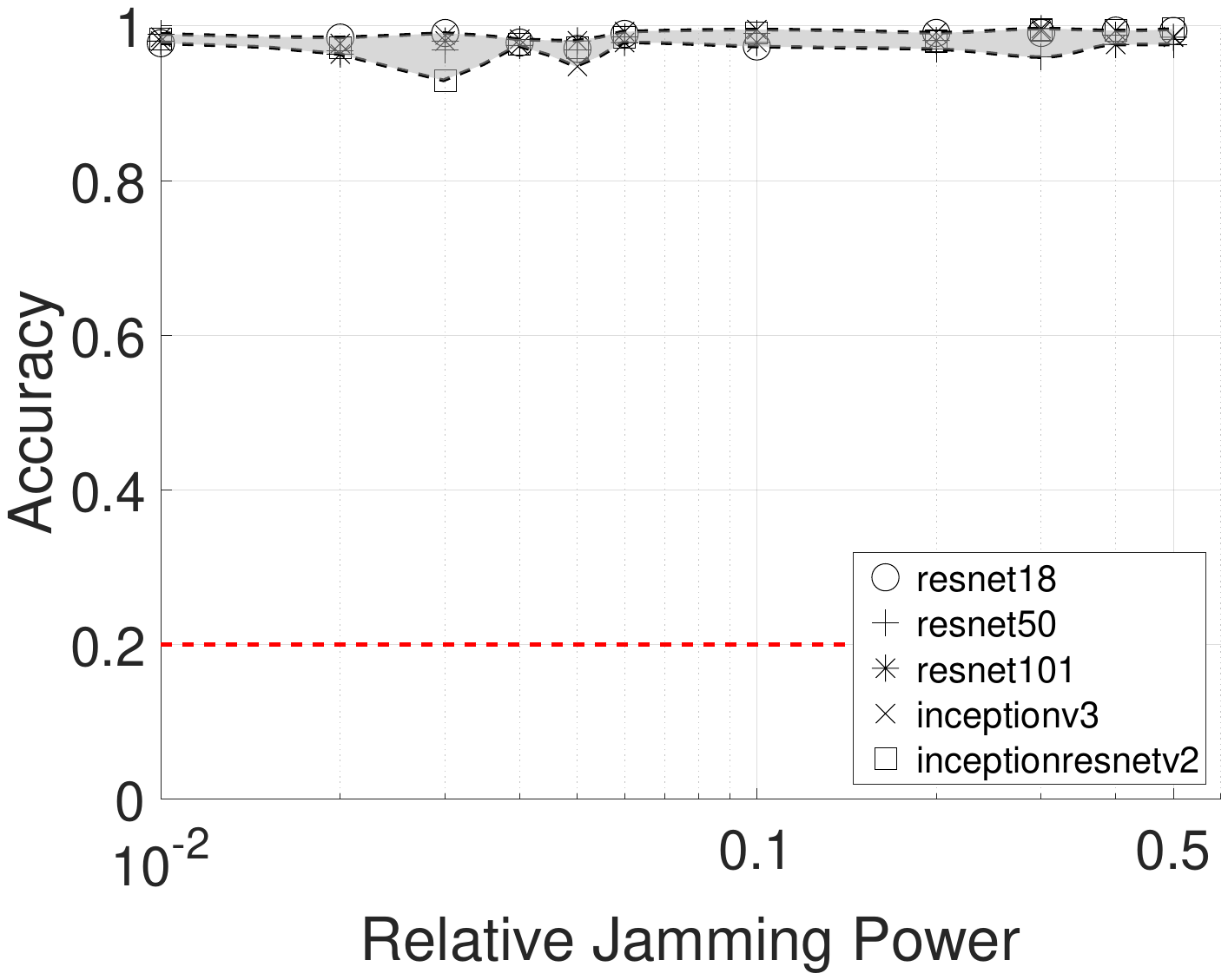}
        \caption{}
    \end{subfigure}    
    \begin{subfigure}{\columnwidth}
        \includegraphics[width=\columnwidth, trim={10mm 22mm 20mm 0}, clip]{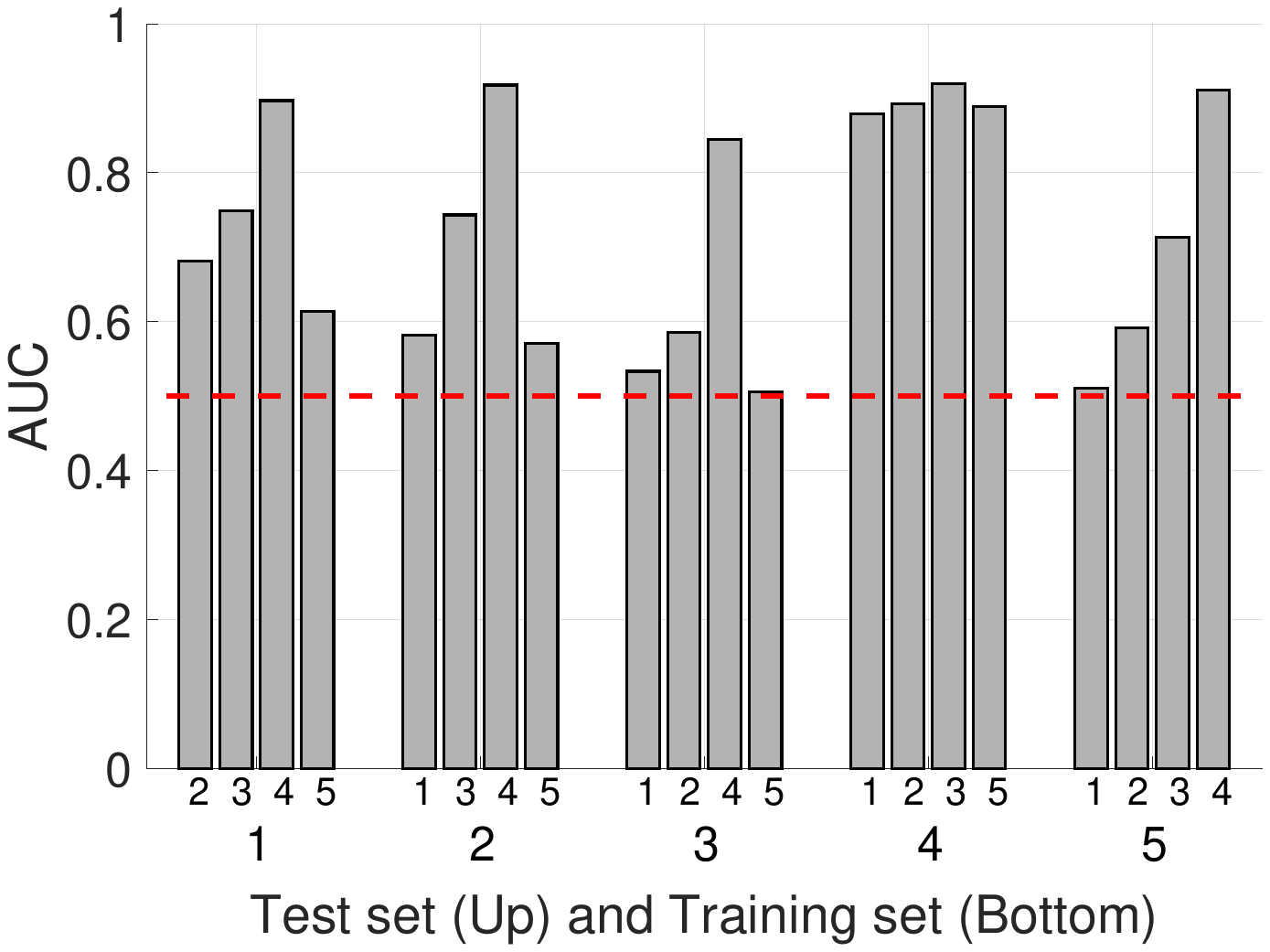}
        \caption{}
    \end{subfigure}
    \caption{(a) k-anonymity. Accuracy as a function of the Relative Jamming Power (RJP) when the training set contains all the data characterized by the same RJP. (b) T-anonymity. Area Under the Curve (AUC) as a function of the transmitter IDs. We challenged a model trained on all the available samples (jammed and not-jammed) from a specific transmitter (bottom labels) against the remainder pool of transmitters one by one (upper labels). The dashed red line represents the random guess for both the subfigures.}
    \label{fig:adv_know}
\end{figure}

\textcolor{\chgclr}{
\section{Adaptive Adversary}
}
\label{sec:adv_knowledge}
In this section, we consider that the adversary eavesdrops and collects both jammed and not-jammed samples, and uses them to train the \ac{DL} model to be used in the subsequent testing phase. 
\\
{\bf k-anonymity.} We first consider k-anonymity, i.e., a pool of $k$ transmitters that want to stay anonymous against the adversary. Recalling the cable set-up from Fig.~\ref{fig:hw_setup}, we assume two scenarios: (i) the adversary collects data samples from a specific \ac{RJP} level and is challenged against the same configuration, and (ii) it collects samples from any available \ac{RJP} and it is challenged to recognize the pool of the transmitters. For the first scenario, i.e., fixed \ac{RJP}, we report the results in Fig.~\ref{fig:adv_know}(a). The adversary is always able to identify the transmitter with high accuracy, i.e., not less than 0.9. As for the second scenario, i.e., the adversary collects samples from any \ac{RJP}, we report only analytical values---being consistent with the ones of the first scenario: we experienced an average accuracy value of 0.95 with a minimum of 0.91. As for the radio set-up (recall Fig.~\ref{fig:hw_setup}), we performed multiple runs of training and testing while shuffling the dataset, and we never observed an accuracy of less than 0.96 (average of 0.98).
\\
{\bf T-anonymity.} We now focus on the scenario involving T-anonymity, i.e., a target device wants to stay anonymous. Our analysis involves all the transmitters. Therefore, we iteratively trained a new model on all the data associated with each transmitter (with and without the jammer), and then, tested against the pool of remaining transmitters (one by one). Figure~\ref{fig:adv_know}(b) shows the AUC associated with the performance of the autoencoders, assuming the different combinations of transmitters considered for training (lower IDs on the x-axis) and testing (upper IDs on the x-axis). The dashed red line highlights the random guess (AUC equal to 0.5). First, we observe that transmitter 4 features the highest distinguishability independently of the considered case, i.e., AUC between 0.85 and 0.92. While we highlight that this is not an outstanding performance for RFF, we also note that the other transmitters perform even worse, on average. While comparing these results with the previous ones, we can conclude that adding jammed samples to the training dataset does not give to the adversary any significant advantage when performing a target attack.
\\
\textcolor{\chgclr}{{\bf Summary.} When considering k-anonymity, an adversary who has the possibility to train on noisy samples for all the possible transmitters can keep classification accuracy high, bypassing the deployment of \sol. However, when considering T-anonymity, additional knowledge does not help the adversary to keep classification accuracy high, confirming the effectiveness of \sol. }

\section{Related Work}
\label{sec:related_work}
Little research has focused on the protection of RF devices from stealthy RFF carried out by a passive adversary. The most relevant work along our line of reasoning is the contribution by Abanto et al.~\cite{abanto2020_macs}, which proposes devise a system to protect against \ac{RFF} solutions using non-linear \ac{CSI} phase errors to identify devices (see~\cite{hua2018accurate} and~\cite{liu2019_icc}). They propose injecting a random phase correction into the signal emitted by the devices so as to reduce the accuracy of such \ac{RFF} systems. Although their solution reports outstanding performance, we argue that there are several problems connected with its usage and validity against modern RFF systems. First, the cited work injects a random phase correction to the transmitted signal by distorting the I-Q samples at the physical layer of the communication stack. Thus, the solution requires one to have access to the signal at the physical layer---a feature that is usually not achieved by commercial-off-the-shelf-devices. Moreover, the authors consider \ac{CSI}-based RFF approaches, which are shown to be reliable for RFF only with WiFi devices. As of today, \ac{DL}-based \ac{RFF} solutions using images are state-of-the-art, as they can infer features characterizing RF systems more effectively, even in the presence of various noise sources. Moreover, the authors do not consider \ac{DL}-based \ac{RFF} systems, and thus, it is impossible to know if their solution is robust also against these approaches. Conversely, \sol\ does not require any modification to the transmitter, as it can be achieved through an external jammer independent of the device to be protected. Moreover, \sol\ protects against RFF even when such systems use state-of-the-art \ac{DL}-based approaches, such as \acp{CNN} and autoencoders.

At the same time, it is worth noting that many contributions in recent years investigated methods to increase the robustness of RFF in the presence of real-world phenomena that could potentially decrease its robustness. If controllable, such phenomena could be used on purpose to protect against RFF.
In the early phases of \ac{RFF}, researchers used feature engineering techniques to devise custom solutions based on \ac{ML} and \ac{DL} techniques characterized by very high accuracy~\cite{riyaz2018deep,sankhe2019oracle,merchant2018deep,yu2019robust,ding2018specific}. 
As of today, although efforts to increase the robustness of RFF further are still significant (see, e.g., recent works such as~\cite{gul2022fine,soltani2020more,al2021deeplora}), researchers are also exploring other facets of the RFF problem in view of its deployment in the real world~\cite{alhazbi2023challenges}. One of the most pressing challenges is the robustness of RFF to challenging channel conditions, as investigated, e.g., by the authors in~\cite{shawabka2020_infocom} and~\cite{pan2024_wlet}. Compared to these works, our solution intentionally decreases the quality of the channel to degrade the accuracy and robustness of RFF approaches while not affecting communication.
Another issue is the light dependence of the RFF model on the specific receiver device. This issue is mentioned by the authors in~\cite{hanna2022wisig}, but it is not exploitable to increase robustness against tracking via RFF since an adversary can always use the same RX device.
Similarly, the authors in~\cite{alhazbi2023dayaftertomorrow} discovered that power-recycling of radios can also affect \ac{RFF} negatively. The authors proposed a pre-processing technique to mitigate the power cycling effect, so this effect is not usable to protect against RFF. Other aspects have been recently found to have an impact on RFF, such as temperature~\cite{gu2024_tosn} and hardware warm-up~\cite{elmaghbub2024_wisec}. In \cite{shen2021radio} the authors investigated the impacts of \ac{CFO} on \ac{RFF}, but also proposed a solution to keep the classification accuracy high. Thus, this method is also not suitable for protecting against RFF.
At the same time, works on the detection of jamming through features of I-Q samples are also relevant to this work due to our proposed solution. 
In this context, Sciancalepore \emph{et. al.}\cite{sciancalepore2023jamming} proposed a technique to detect jamming in a low bit-error rate regime for indoor mobile scenarios. The authors employed sparse autoencoders to distinguish between benign and jamming-affected signals by converting the physical layer (I-Q) signals into images. Another solution toward the early detection of jamming is addressed by Alhazbi \emph{et. al.}\cite{alhazbi2023ccnc}. The authors introduced a \ac{DL}-based system for early detection and identification of jamming in wireless networks. By analyzing I-Q samples from signals, the solution accurately identifies various jamming types, such as Gaussian noise and tone-jamming, before they disrupt communications. This proactive approach enhances network resilience and enables timely countermeasures, significantly improving the robustness of wireless systems against jamming attacks. However, the focus of these papers is on jamming detection and not on the RFF of the devices in the presence of jamming. Thus, they do not directly apply to our scenario.

Finally, we notice that there are solutions in the literature to recover the bits of a communication link under jamming conditions, mostly known as techniques for jamming cancellation and mitigation, e.g., the recent solution in~\cite{nguyen2023_wisec}. However, such solutions recover the bits of the communication following a jamming attack, although the corresponding IQ samples at the physical layer have been displaced from their original position in the IQ plane due to jamming. Since the radio fingerprint is well represented by the displacement of the IQ samples in the IQ plane, as per Fig.~\ref{fig:snr_computation} \textcolor{\chgclr}{from Sec.~\ref{sec:measurement}}, such solutions are not applicable to move back the IQ samples in their original position---this is not the scope of jamming cancellation and mitigation solutions. 
Overall, given the current state of the art, there is no solution to the reconstruction of the radio fingerprint after it has been removed. As soon as the IQ samples are re-arranged by either the radio noise or by collisions with another signal as per our solution, the information related to the fingerprint is permanently removed, and there is no way to reconstruct it.

In summary, to the best of our knowledge, this is the first contribution to specifically focus on the protection of anonymity of wireless communications against state-of-the-art \ac{DL}-based \ac{RFF} approaches. Also, it is the first solution that can work out-of-the-box independently from the communication technology used by the device to be protected, demonstrating significant potential for real-world privacy-preserving deployments.

\section{Conclusion}
\label{sec:conclusion}
In this work, we proposed the concept of jamming-based fingerprint sanitization as a practical approach to preserving the privacy and anonymity of wireless transmitters. As distinctive features, our proposed solution, namely \sol, enables fingerprint sanitization without compromising communication quality, providing k-anonymity and T-anonymity for the transmitting devices while maintaining data integrity. Also, it does not require access or hardware modifications to wireless devices, being lightweight and non-invasive.
To investigate the effectiveness of our approach, we conducted an extensive measurement campaign that involved both wired and wireless measures using \aclp{SDR}. The results demonstrate that deliberate low-power jamming can effectively hide the unique hardware fingerprints required for radio fingerprinting techniques, making it more difficult (if not impossible) for the adversary to identify transmitters in an unauthorized fashion.
This approach offers a viable solution to protect transmitters from being identified by malicious eavesdroppers who employ state-of-the-art deep learning-based identification techniques.
Future works involve further analysis of our approach to investigate its robustness against an adversary who can re-train a model for the transmitter with various levels of jamming.


\bibliographystyle{IEEEtran}
\balance
\bibliography{main}

\end{document}